\documentclass{aa}  

\usepackage{graphicx}
\usepackage{booktabs}
\usepackage[switch]{lineno}
\usepackage[linesnumbered,ruled,vlined]{algorithm2e}
\usepackage{xcolor, amsmath}
\usepackage{txfonts}
\usepackage[colorlinks=true, linkcolor=blue, citecolor=blue, urlcolor=blue]{hyperref}
\begin{document}

   \title{The Star Formation Factory revisited }

   \subtitle{I. The impact of metallicity on collapsing star-forming clouds}

   \author{S. Jim\'enez\inst{1}, D. Kom\'anek\inst{1,2}, R. Wünsch\inst{1}, J. Palouš\inst{1}, S. Ehlerov\'a\inst{1}, S. Mart\'{\i}nez-Gonz\'alez\inst{3}, \and A. Srbljanovi\'c\inst{1} \fnmsep
          }

   \institute{Astronomical Institute of the Czech Academy of Sciences,  Bo\v{c}n\'\i\ II 1401/1, 141 00 Praha 4, Czech Republic\\
              \email{santiago@asu.cas.cz}
         \and
             Faculty of Mathematics and Physics, Charles University, V Hole\v{s}ovi\v{c}k\'ach 2, 180 00 Praha 8, Czech Republic
         \and
             Instituto Nacional de Astrof\'\i sica \'Optica y Electr\'onica, AP 51, 72000 Puebla, M\'exico
            }

   \date{Received XX, 2025; accepted XX, 2025}

 
\abstract
{Stellar feedback regulates star formation and shapes the interstellar medium, yet its role during the collapse of molecular clouds remains uncertain over a wide range of initial conditions.}
{We explore how stellar winds and supernovae influence star formation in collapsing gas clouds that span a broad parameter space in mass, size, and metallicity.}
{Using a one-dimensional numerical model, we follow the evolution of feedback-driven bubbles produced by embedded clusters, incorporating time-dependent energy and mass injection, self-gravity, integrated cloud collapse, radiative cooling, shell instabilities, and triggered star formation. Our treatment of gas cooling in the hot bubble explicitly accounts for heat transfer across the bubble–shell interface.}
{We find that metallicity acts as a key regulator of feedback, comparable in importance to cloud mass and radius. In low-metallicity clouds, reduced radiative cooling is offset by weaker stellar winds, leading to prolonged star formation and higher efficiencies. Across a substantial portion of parameter space, the expanding shell undergoes a stalling phase that further enhances the star formation efficiency, an outcome that is not observed at higher metallicities.}
{Our results suggest that the diverse properties of star clusters across cosmic time may arise from the metallicity-dependent interplay between stellar feedback and gas cooling.}


  \titlerunning{The Star Formation Factory I: Impact of metallicity}
\authorrunning{S. Jiménez et al.}
   
   \maketitle
\modulolinenumbers[10]          
\setlength\linenumbersep{6pt}   
%

\section{Introduction}
Most stars in the Universe are born within giant molecular clouds (GMCs), where gravitational collapse drives star formation until the available gas is either consumed or expelled by stellar feedback \citep[e.g.][]{2019Natur.569..519K,2020MNRAS.493.2872C, 2021ApJ...914...90L, 2021MNRAS.506.2199G,2022MNRAS.517.1313M, 2024ApJ...967L..28M}. The fraction of the original cloud mass that is converted to stars, the star formation efficiency (SFE), depends on the balance between competing forces, gravity and the cloud ram pressure, which promotes collapse, and stellar feedback, that disperses and clears the surrounding medium.

In recent years, considerable progress has been made in understanding star formation and stellar feedback on a wide range of scales, from entire galaxies \citep[e.g.][]{2018MNRAS.477.2716K, 2020ApJ...903L..34K, 2022ApJ...936..137O,2023ApJ...956..118P}, down to individual molecular clouds \citep[e.g.][]{2019ARA&A..57..227K, 2021MNRAS.506.2199G}.   Observations robustly show that star formation is inefficient in GMCs ($\sim 1-10$ \%, \citealt{2017ApJ...846...71L, 2018ApJ...861L..18U}). On the theoretical side, numerical simulations now include multiple feedback processes simultaneously \citep[e.g.][]{2019ApJ...879L..18L, 2022MNRAS.512..216G,2023MNRAS.521.5686K,2024MNRAS.534..215F, 2024A&A...691A.231D, 2024MNRAS.530..645L, 2025MNRAS.540.1462R}, which is fundamental to properly understanding the integrated impact of stellar feedback. Nevertheless, and despite its importance, there are currently  still many open questions regarding the relative importance and efficiencies of the different feedback mechanisms \citep[e.g.][]{2019ARA&A..57..227K, 2025ApJ...989...42L}.

Winds and radiation from young massive stars are widely viewed as the main mechanisms that halt star formation in GMCs within short timescales  \citep[e.g.][]{2010ApJ...709..191M,2011ApJ...738...34P,2019MNRAS.486.5263M, 2019ARA&A..57..227K}. This could explain observations showing a de-correlation on small scales between young stellar regions and molecular gas tracers \citep[e.g.][]{2019MNRAS.490.4648H, 2019Natur.569..519K,2021ApJ...918...13S}. Indeed, the combined mechanical and radiative feedback from individual stars merges to create large thermal pressures, leading to cluster-scale winds \citep[e.g.][]{1975ApJ...200L.107C,1985Natur.317...44C,2000ApJ...536..896C, 2004ApJ...610..226S, 2017MNRAS.470.4453R}. This can very efficiently evacuate the remaining gas from the clouds, as suggested by observations, which indicate that the gas-embedded phase of star cluster formation is short ($\lesssim 10$ Myr, e.g. \citealt{2018MNRAS.481.1016G, 2019MNRAS.483.4707G,2022MNRAS.516.4612T,2025arXiv250701508R}).
\begin{figure*}[htb]
\centering
	\includegraphics[width=1.6\columnwidth]{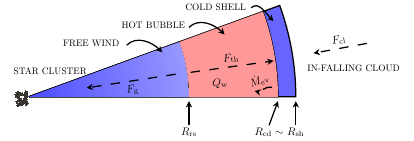}
\caption{Schematic of the model setup of this work. We model the evolution of feedback-driven bubbles, formed by the winds and supernovae from the star cluster forming at the center. The scheme presents a cone of the otherwise spherically symmetric bubble structure, showing the location of the reverse ($R_{\rm{rs}}$) and forward shocks ($R_{\rm{sh}}$), and the contact discontinuity ($R_{\rm{cd}}$) which separates shocked wind from ambient swept-up gas. Furthermore, the dashed arrows indicate the directions of the different forces that determine the evolution of the shell radius: the inward gravitational force ($F_{\rm{g}}$) and ram pressure from the cloud ($F_{\rm{cl}}$), and the outward force produced by the thermal energy of the hot bubble ($F_{\rm{th}}$). This term is calculated by considering radiative cooling within the bubble ($Q_{ \rm{w}}$), which includes the effect of mass evaporation from the cold shell. The evaporated material ($\dot{M}_{\rm{ev}}$) mass-loads the bubble interior, thus modifying its density and temperature structure and, consequently, the cooling rate. In addition, our model setup includes star formation in the shell (not included for clarity of the diagram). See the text for details.}
\label{scheme}
\end{figure*}

However, in some environments stellar feedback might be greatly reduced or suppressed, allowing star formation to proceed over longer timescales thus enabling chemical enrichment within the star-forming cloud and the forming stars \citep[e.g.][]{2013A&A...552A.121K,2014ApJ...792..105P,2017MNRAS.470..977L,Wunschetal2017,2017MNRAS.465.1375S,2018MNRAS.476.5341F,2019MNRAS.486.1146H,2021MNRAS.505.4669J, 2021ApJ...922L...3L, 2022MNRAS.513.2360J, 2023ApJ...958..149J}. In addition, stellar feedback also triggers star formation in some cases \citep[e.g.][]{2009MNRAS.398.1537D,2017Ap&SS.362..183R,2023MNRAS.520.5600D}. For instance, \cite{2003A&A...411..397T} proposed a scenario for the formation of massive star clusters, in which feedback from a young stellar population drives the expansion of a shell that eventually becomes gravitationally unstable, triggering further star formation in the shell. \cite{2003A&A...411..397T} showed that, under certain conditions, the feedback can balance gravity and infall, stalling the shell while enabling continuous star formation. This mechanism has been suggested as a pathway to form massive star clusters.

Despite significant advances in recent years, the role of stellar feedback in regulating star formation remains highly debated. This is in part due to the interplay of multiple feedback processes, such as stellar winds, ionizing radiation, and supernova explosions, which are challenging to model self-consistently within a single simulation setup. Moreover, the physical conditions of star-forming regions vary widely across environments, from nearby molecular clouds to high-redshift galaxies, complicating efforts to assess the relative importance of different feedback channels and their influence on star formation efficiency and timescales \citep[e.g.][]{2024A&A...690A..72F,2024A&A...681A..28A}. This is particularly the case for low-metallicity environments, where observational constraints and theoretical models remain limited \citep[e.g.][]{2014MNRAS.442.3112P,2023ApJ...958..149J}. Metallicity directly affects cooling, dust content, radiation coupling, and wind power, all of which modulate how effectively young stars can disrupt their natal clouds. Understanding feedback under such conditions is therefore essential to bridge the gap between local, nearby star-forming regions and those in the early Universe.  

Furthermore, the unprecedented sensitivity and resolution of the \textit{James Webb Space Telescope} (\textit{JWST}) are now revealing massive, compact, and metal-poor star-forming complexes at high redshift \citep[e.g.][]{2023A&A...678A.173V,2024Natur.636..332M, 2025MNRAS.540.2176C, 2025ApJ...983L..22K, 2023ApJ...957...77P, 2024ApJ...976..166P, 2024Natur.632..513A}, providing new constraints on the physical processes that regulate their evolution. These observations highlight the need for theoretical models capable of predicting how feedback operates in such extreme, low-metallicity conditions, where traditional feedback-regulated star formation paradigms may break down \citep[e.g.][]{2025ApJ...990..173C}.

To explore these processes in detail, here we study how winds and supernovae feedback impact infalling star-forming molecular clouds by means of 1D calculations that include several physical processes. Our calculations include time-dependent energy injection by the central cluster, cluster gravity and the shell self-gravity, and a Larson–Penston-like inflow. Our models build upon the classical structure of feedback-driven bubbles \citep[e.g.][]{1975ApJ...200L.107C,Weaver1977}, incorporating the hot shocked wind region, the swept-up shell, and their interaction through thermal conduction.

We introduce a novel implementation for hot-bubble cooling that accounts for heat transfer across the hot–cold interface, which allows for gravitational fragmentation of the swept-up shell, enabling secondary star formation. Finally, we further explore the role of metallicity, which affects both the gas cooling and the stellar wind mechanical power.

Our aim is twofold. With our setup, we will first establish the conditions under which the model proposed by \cite{2003A&A...411..397T} applies. Second, we will study more generally when winds and supernovae suppress or trigger star formation. By covering a wide parameter space in cloud mass, size, and metallicity, our study provides new insights into how feedback regulates star cluster formation and sheds light on the origin of very old, low-metallicity globular clusters in the early Universe.

This paper is organized as follows. Section \ref{sec_setup} describes in detail the model setup, including the physics included in our calculations and the set of models selected for our study. Section \ref{sec_Results} presents and discusses the general outcomes of our calculations. Section \ref{SFF_section} discusses the standing shell solution and its properties and dependence on the input parameters. Section \ref{sec_Discussion} discusses our main results and the model limitations and section \ref{sec_Conclusions} presents our summary and main conclusions.

\section{Model setup} \label{sec_setup}
Fig. \ref{scheme} presents a schematic of our model setup.  We model the evolution of feedback-driven bubbles (section \ref{sec_struct_shells}), which assumes a spherically-symmetric collapsing star-forming cloud (section \ref{cloud_density}), with a point-like stellar cluster located at the cloud core forming with a given star formation rate (section \ref{star_formation_rate}). Furthermore, we consider secondary star formation which occurs due to the gravitational fragmentation of the shell (section \ref{add_star_formation}). The stars inject mechanical feedback by stellar winds and supernovae (section \ref{energy_input_pro}) thus creating a bubble-driven shell. Our model includes the free wind region extending from the center to the inner reverse shock, $R_{\rm{rs}}$, the shocked wind region between $R_{\rm{rs}}$ and the contact discontinuity $R_{\rm{cd}}$, and the shocked ambient gas shell between $R_{\rm{cd}}$ and the forward shock $R_{\rm{sh}}$, which we assume to be infinitesimally thin (i.e. $R_{\rm{cd}} \sim R_{\rm{sh}}$ ; see section \ref{shell_evolution}). We also include the gravitational force exerted by the central star cluster, the self-gravity of the shell, and radiative cooling in the hot bubble (section \ref{energy_equation}). Specifically, our calculation of the cooling rate takes into account heat conduction (and thus the mass evaporation) between the cold shell and the hot bubble (Appendix \ref{app_evapora}).  Finally, section \ref{initial conditions} describes the initial conditions of our calculations and defines the parameter space that was considered for this work. 

\subsection{The structure of feedback-driven shells}\label{sec_struct_shells}
The typical structure of feedback-driven bubbles can be understood in terms of a sequence of zones (see Fig. \ref{scheme}) and characteristic timescales, each corresponding to a distinct stage in its evolution \citep[e.g.][]{1975ApJ...200L.107C,Weaver1977,bisno1}. From the inside out, the bubble consists first of a freely expanding wind, extending outwards until it is terminated by a reverse shock at radius $R_{\rm{rs}}$. At this point, the wind is shock-heated to high temperatures ($\sim 10^6$–$10^7$ K) and forms the hot bubble, going from $R_{\rm{rs}}$ up to a contact discontinuity ($R_{\rm{cd}}$). The outermost layer of the bubble, which starts at $R_{\rm{cd}}$, is composed of the swept-up ambient gas, processed by the leading shock at radius $R_{\rm{sh}}$, which defines the boundary between the bubble and the surrounding gas cloud. As shown by \cite{Weaver1977}, the outer layer cools quickly, forming a dense, cold and thin shell, but the inner rarefied bubble of shock-heated wind ejecta (hereafter, the hot bubble) can sustain the expansion of the shell for longer timescales. Thus, as shown in Fig. \ref{scheme}, here we will assume that the shell of swept-up material is thin and thus $R_{\rm{cd}} \sim R_{\rm{sh}}$. Furthermore, in the interface between the hot bubble and the cold shell, heat conduction can mass-load the bubble interior thus altering the cooling rate. It is important to include this process into the calculations as it impacts the structure and dynamical evolution of feedback-driven bubbles \citep[e.g.][]{2019MNRAS.490.1961E, 2025ApJ...989...42L}.

\subsection{The star-forming gas cloud} \label{cloud_density}
Here we consider a star-forming cloud given by a power-law density distribution with an inner core:
\begin{equation}\label{gas_cloud}
\rho_{\rm{cl}}=\left\{
	\begin{array}{ll}
		\rho_{\rm{c}},  & \mbox{if }  r \leq R_{\rm{c}},\\
		\rho_{\rm{c}}\left(\frac{r}{R_{\rm{c}}} \right)^{-2} & \mbox{if } R_{\rm{c}} < r \leq R_{\rm{cl}}.\\
	\end{array}
\right.
\end{equation}
where $\rho_{\rm{c}}$ and $R_{\rm{c}}$ are the core density and radius, $\alpha$ is the power-law index, and $R_{\rm{cl}}$ is the cloud radius. The adopted density profile corresponds to the asymptotic structure for approximately isothermal, self-gravitating clouds. Observed clouds \citep[e.g.][]{2021MNRAS.502.4963G} and numerical simulations have been shown to follow approximately such density profile \citep[e.g.][]{2018MNRAS.473.4220L,2025MNRAS.541.3880Z}.  Furthermore, we also adopt a Larson-Penston velocity field, which is the solution for collapsing isothermal clouds \citep{1969MNRAS.145..271L, 1969MNRAS.144..425P}, given by:
\begin{equation}\label{gas_cloud_vel}
v_{\rm{cl}}=\left\{
	\begin{array}{ll}
		\frac{R}{R_{\rm{c}}} v_{\rm{cl,m}},  & \mbox{if }  r \leq R_{\rm{c}},\\
		v_{\rm{cl,m}} & \mbox{if } R_{\rm{c}} < r \leq R_{\rm{cl}},\\
	\end{array}
\right.
\end{equation}
where $v_{\rm{cl,m}}=3.3 c_{\rm{s}}$ is the maximum infalling velocity and $c_{\rm{s}}$ is the sound speed in the cloud. Note that, since collapse is included in the calculations, both $R_{\rm{c}}$ and $R_{\rm{cl}}$ are functions of time. Thus, hereafter the subscript ``0'' will denote the initial values of all variables at the initial time $t_{\mathrm{ini}}$ (see section \ref{initial conditions} for a definition for $t_{\mathrm{ini}}$), that is, for this case $R_{\rm{c,0}}=R_{\rm{c}}(t_{\mathrm{ini}})$ and $R_{\rm{cl,0}}=R_{\rm{cl}}(t_{\mathrm{ini}})$, respectively. Since our aim is to isolate the impact of stellar feedback in collapsing star-forming clouds, we adopt a core radius that is a fixed fraction of the cloud radius ($R_{\rm{c}}=0.2 R_{\rm{cl}}$) in order to preserve a self-similar density structure across the explored parameter space. This choice ensures that variations in the results primarily reflect differences in global cloud properties (e.g. mass and radius), rather than changes in the internal structure of the cloud.

The initial value of the core density $\rho_{\rm{c}}$ is calculated from the cloud initial gas mass $M_{\rm{g,0}}$ and radius $R_{\rm{cl,0}}$, which are part of our input parameters.  We assume that the cloud is in thermal equilibrium with the low-density ambient medium ($r> R_{\rm{cl}}$), but this does not affect our results as we stop all models when the shell reaches the cloud boundary. 
\begin{figure}[htb]
\centering
	\includegraphics[width=\columnwidth]{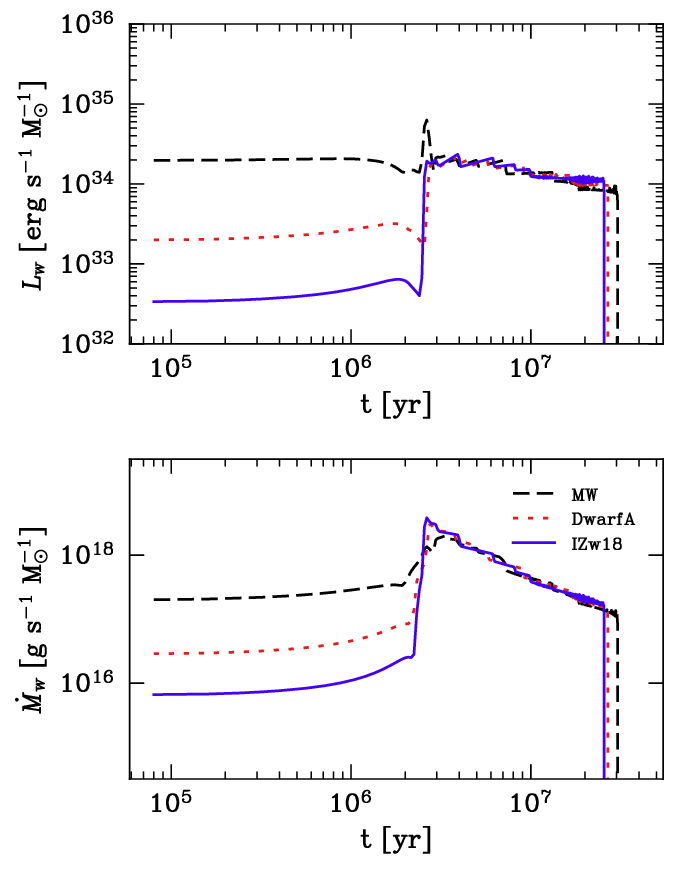}
\caption{Evolution of the mechanical power (top) and mass input rate (bottom) per unit solar mass of a stellar population with MW (dashed lines), dwarfA (dotted lines) and IZw18 (solid lines) metallicities as obtained using the Bonn Optimized Stellar Tracks \citep[BoOST;][]{BoOST2022}.}
\label{inputPower_plot}
\end{figure}
\subsection{Star formation rate in the central core}\label{star_formation_rate}
We model the star cluster forming at the center of the cloud (with mass $M_{\rm{sc}}$) as a continuous process given by a star formation rate \citep[e.g.][]{2007ApJ...654..304K, 2012ApJ...745...69K,2016ApJ...831...73V, 2016ApJ...833..229L, 2018ApJ...854...16E,2023ApJ...945L..19S}: 
\begin{equation} \label{core_SFR}
\frac{\rm{d} M_{\rm{sc,c}}}{\rm{d}t}= \epsilon_{\rm{ff}} \frac{M_{\rm{c}}}{t_{\rm{ff}}},
\end{equation}
where $M_{\rm{c}}$ ($M_{\rm c} \sim 0.1M_{\mathrm{gas}}$) is the gas mass within the core ($R \leq R_{0}$), $\epsilon_{\rm{ff}}$ is the star formation efficiency per free-fall time, and
\begin{equation}
t_{\rm{ff}}= \sqrt{\frac{3\pi}{32G \rho_{\rm{c}}}}
\end{equation}
is the core free-fall time. The core mass $M_{\rm c}$ is calculated at every time step from mass conservation, by taking into account the mass swept-up by the shell and the mass depleted by star formation. 

The dependence of $\epsilon_{\rm{ff}}$ on metallicity remains not fully understood. Some studies report a decrease in $\epsilon_{\rm{ff}}$ with metallicity, whereas others find it to be nearly independent of it \citep[e.g.][]{2011MNRAS.415.3439D,2024OJAp....7E.114P}. In our models, $\epsilon_{\rm{ff}}$ is a free parameter and we adopt $\epsilon_{\rm{ff}} \in \left(0.01, 0.1\right)$ as a representative range \citep[e.g.][]{2021ApJ...912L..19P,2025MNRAS.541.3880Z}. However, for two low-metallicity cases, we explore slightly higher efficiencies to examine the impact of this parameter (see Table \ref{tab:sim_params}).

Star formation in the core is allowed only while $t \leq t_{\rm{ff,0}}$ as this is also approximately the time required for the shell to overrun the core of the density distribution. Our calculations assume that $M_{\rm{sc,0}}=100$ M$_{\odot}$. Since the density and other cloud properties evolve with time, the prescription given by Equation \ref{core_SFR} naturally yields a time-dependent SFR.

\subsection{Secondary star formation}\label{add_star_formation}
Several hydrodynamical and gravitational instabilities can develop in a cold, dense, expanding shell. Thin-shell (Vishniac-type; e.g., \citealt{1983ApJ...274..152V, 1989ApJ...337..917V, 2013A&A...550A..49K}) and Rayleigh–Taylor instabilities can induce nonlinear density structure within the shell. However, the formation of gravitationally bound fragments requires self-gravity to overcome internal pressure and expansion effects. In particular, previous studies of the Vishniac instability \citep[e.g.,][]{1993ApJ...407..207M,2013MNRAS.435.3600P} show that while it efficiently generates density perturbations, it does not by itself ensure the development of gravitationally bound fragments. 

Once the shell accumulates sufficient mass (and thus increases its surface density), self-gravity can overcome internal pressure and expansion, leading to gravitational fragmentation. In this regime, the growth rate of gravitational instabilities sets the timescale for collapse and thus regulates the star formation rate \citep[e.g.][]{1994A&A...290..421W, 1994MNRAS.268..291W, 1994ApJ...427..384E}.

Our star formation prescription therefore explicitly links the formation rate of stellar mass to the peak growth rate of gravitational instability in the shell:

\begin{equation}
\frac{{\rm d} M_{\rm sc,s}}{{\rm d}t} =
\begin{cases}
\omega_{\rm p} M_{\rm sh}, & \text{if } \omega_{\rm p} \ge 0 \text{ and } \lambda/R_{\mathrm{sh}} \le 1, \\
0, & \text{otherwise.}
\end{cases}
\label{cond_fragmentation}
\end{equation}
where
\begin{equation}
\omega_{p}=-\frac{3U}{R_{\mathrm{sh}}}+\left[\frac{U^2}{R_{\mathrm{sh}}^2}+ \left( \frac{\pi G \Sigma_{\rm{sh}}}{ c_{\rm{s}}}\right)^2 \right]^{1/2}
\label{star_formation_recipe}
\end{equation}

is the maximum growth rate of gravitational perturbations (with wavelength $\lambda$) in the expanding shell \citep{Elmegreen1994, richard2001}. In these equations, $R_{\rm{sh}}$, $U$, $M_{\rm{sh}}$, and $\Sigma_{\rm{sh}}$ are the shell position, velocity, mass, and surface density, respectively. The term $c_{\rm{s}}=1$ km s$^{-1}$ is the sound speed in the shell. 

Note that star formation, as given by equation \ref{cond_fragmentation}, only occurs if two criteria are met by the expanding shell: $\omega_p$ should be positive and the perturbation wavelength ($\lambda$) should not be larger than the shell size. The second condition can be written as \citep{sonia1997}:
\begin{equation}
    \frac{2 c_{s}^{2}}{G \Sigma_{\rm{sh}}R_{\rm{sh}}}<1.
\label{star_formation_recipe2}
\end{equation}

Although equation \ref{star_formation_recipe} is derived from a linear stability analysis, we do not assume that the shell remains in the linear regime. Rather, $\omega_p ^{-1}$ is used as an effective collapse timescale that characterizes the rate at which self-gravity can convert shell gas into bound fragments once sufficient compression and surface density have been achieved.

We further assume, as \cite{2003A&A...411..397T}, that the newly formed stars free fall towards the central region given their negligible cross-section, thus effectively increasing the value of $M_{\mathrm{sc}}$.

\subsection{Energy and mass input rates}\label{energy_input_pro}
Stars form in our calculations following the initial mass function (IMF) by \cite{Maschberger2013}, with lower and upper mass cutoffs of 0.01 M$_{\odot}$ and 120 M$_{\odot}$, respectively. 

In this work, we restrict our analysis to mechanical feedback from massive stars, specifically stellar winds and supernova explosions, in order to assess the conditions under which the model of \cite{2003A&A...411..397T} is applicable. In future work, we will include additional feedback channels, such as radiation feedback,  to evaluate their relative contributions and to better constrain the overall impact of the different mechanisms.  

The feedback parameters from individual stellar populations (with contributions from all stars with initial masses $m \geq 9$ M$_{\odot}$), are obtained from the simple population synthesis code SYNSTARS using the Bonn Optimized Stellar Tracks \citep[BoOST;][]{BoOST2022}, where the \cite{kk21} prescription for $v_{\inf}/v_{\rm{esc}}$ was employed for OB stars, and \cite{sv20} for the Wolf-Rayet (WR) stars. Furthermore, the terminal velocity of the wind is set to a constant value of $30$ km s$^{-1}$ for red supergiants \citep{2010ApJ...723.1210A} and 200 km s$^{-1}$ for luminous blue variable stars \citep{2018A&A...619A..54V}.

In order to study the importance of metallicity for our models, we have selected the MW ($Z=0.0088$, and [Fe/H]$\sim$ 0.0), dwarfA ($Z=0.00105$, and [Fe/H]$\sim -1.0$) and the IZw18 ($Z=0.00021$, and [Fe/H]$\sim -1.7$) metallicity tracks in BoOST (see also \citealt{IZw18_2015}). Fig. \ref{inputPower_plot} presents the mechanical power ($L_{\rm{w}}$, top panel) and mass input rate ($\dot{M}_{\rm{w}}$, bottom) per unit solar mass as a function of time, for MW (dashed lines), dwarfA (dotted lines) and IZw18 (solid lines) metallicities, respectively.

Stellar evolution could be an important factor as star formation might last long for many of our models. To account for the time dependence of stellar feedback, we divide the elapsed time since the onset of star formation into $N_{\rm{sb}}$ discrete intervals at each simulation time t:
\begin{equation}
\Delta t=(t-t_{\mathrm{ini}})/N_{\rm{sb}}.
\end{equation}
The total energy input rate is then calculated by adding up the contribution from each interval:
\begin{equation}\label{energy_evolving}
L_{\rm{w}} \sim \sum_{i=1}^{N_{\rm{sb}}} L_i \left(t-t_i \right)
\end{equation}
with:
\begin{equation}
t_i=t_0+i \Delta t,
\end{equation}
and:
\begin{equation}\label{eq_Lumi}
L_{\rm{i}} \left(t-t_{\rm{i}} \right) \approx L^{*}\left(t-t_{\rm{i}} \right) \times M_{{\rm{i}}},
\end{equation}
where $L^{*}\left(t-t_{\rm{i}} \right)$ is the interpolated power per unit mass from the stellar template and $M_{\rm{i}}$ is the stellar mass formed during each time interval:
\begin{equation}
M_{\rm{i}}=M_{{\rm{sc}}}\left(t_i\right)-M_{{\rm{sc}}}\left(t_{i-1}\right).
\end{equation}
All our calculations assume a fixed $N_{\rm{sb}}=50$ (i.e., we re-bin at every time step and thus $\Delta t$ changes with time), however, we verified that increasing the number of intervals do not alter the results. 
\subsection{Main equations}\label{shell_evolution}
The shell evolution is dictated by the balance between stellar feedback, inward gravitational force and the ram pressure of the cloud:
\begin{equation}\label{mod1}
\frac{\rm{d}U}{\rm{d}t}=\frac{4 \pi R_{\rm{sh}}^2 \left(P_{\rm{th}}-\rho_{\rm{cl}}v_{\rm{cl}}^2\right)}{M_{\rm{sh}}}-\frac{F_{\rm{g}}}{M_{\rm{sh}}}-\frac{U}{M_{\rm{sh}}}\frac{\rm{d} M_{\rm{sh}}}{\rm{d}t},
\end{equation}

\begin{equation}\label{mod3}
\frac{\rm{d} M_{\rm{sh}}}{\rm{d}t}=4 \pi R_{\rm{sh}}^{2} \rho_{\rm{cl}}\left(U-v_{\rm{cl}} \right) -\frac{\rm{d} M_{\rm{sc,s}}}{\rm{d}t},
\end{equation}
\begin{equation}\label{mod3b}
    \frac{{\rm d} M_{\rm sc}}{{\rm d}t} = \frac{{\rm d} M_{\rm sc,c}}{{\rm d}t} + \frac{{\rm d} M_{\rm sc,s}}{{\rm d}t}
\end{equation}

\begin{equation}\label{mod4}
\frac{\rm{d} R_{\rm{sh}}}{\rm{d}t}=U.
\end{equation}
In Equation \ref{mod1}, $P_{\rm{th}}$ is the thermal pressure and:
\begin{equation}
F_{\rm{g}}=\frac{GM_{\rm{sc}}M_{\rm{sh}}}{R_{\rm{sh}}^2}+\frac{GM_{\rm{sh}}^2}{2R_{\rm{sh}}^2},
\end{equation}
is the gravitational force from the star cluster and the self-gravity of the shell, and G is the gravitational constant. We have omitted the ambient gas pressure term in equation \ref{mod1}, since it is negligible compared to the ram pressure of the infalling cloud. Moreover, all of our simulations stop once the expanding outer shell has completely swept up the cloud. Equation \ref{mod3} describes the evolution of the shell mass and includes mass loss due to star formation. However, we neglect mass loss from evaporation at the shell–bubble interface, as this contribution is small compared to the other terms. The equation \ref{mod3b} specifies that the cluster mass grows through star formation occurring both in the core and in the shell, as discussed in Sections \ref{star_formation_rate} and \ref{add_star_formation}. Finally, note that the cloud density ($\rho_{\rm{cl}}$) is time-dependent, as described in section \ref{cloud_density}.

\subsection{The energy equation}\label{energy_equation}
The thermal pressure in the hot bubble is:
\begin{equation}
    P_{\rm{th}}=(\gamma-1) \frac{3E_{\rm{th}}}{4 \pi(R_{\rm{sh}}^3-R_{\rm{rs}}^3) },
\end{equation}
where $\gamma=5/3$ is the ratio of specific heats, $E_{\rm{th}}$ is the thermal energy and $R_{rs}$ is the reverse shock position. 

The reverse shock position ($R_{\rm{rs}}$) is obtained by solving the implicit equation resulting from the pressure equilibrium at the reverse shock position:
\begin{equation}
    R_{\rm{rs}}^2=F_{\rm{ram}}(R_{\rm{sh}}^3-R_{\rm{rs}}^3)/(2E_{\rm{th}}),
\end{equation}
where $F_{\rm{ram}}$ is the ram pressure of the unshocked stellar wind. 

The thermal energy is calculated as:
\begin{equation} \label{energy_eq}
  \frac{\rm{d}E_{\rm{th}}}{\rm{d}t}  =L_{\rm{w}}- 4\pi R_{\rm{sh}}^2UP_{\rm{th}}-Q_{\rm{w}},
\end{equation}
where 
\begin{equation}
   Q_{\rm{w}}= n_{\rm{H,w}} ^2 \left(\frac{n_{\rm{e,w}}}{n_{\rm{H,w}}}\right)\Lambda \left(T_{\rm{sw}}, Z \right),
\end{equation}
is the energy loss due to radiative cooling in the hot bubble. Here, \( n_{\rm{H,w}} \) and \( n_{\rm{e,w}} \) are the ion and electron number densities, respectively, and \( T_{\rm{sw}} \) is the temperature in the hot bubble. The function \( \Lambda \) denotes the collisional ionization equilibrium (CIE) cooling function. We use the cooling table from \cite{Schure2009} to obtain \( \Lambda(T_{\rm{sw}}, Z) \) and the ratio \( n_{\rm{e,w}}/n_{\rm{H,w}} \), uniformly scaled to the metallicities adopted in our calculations. 
   \begin{figure*}[htb]
\centering
	\includegraphics[width=1.7\columnwidth]{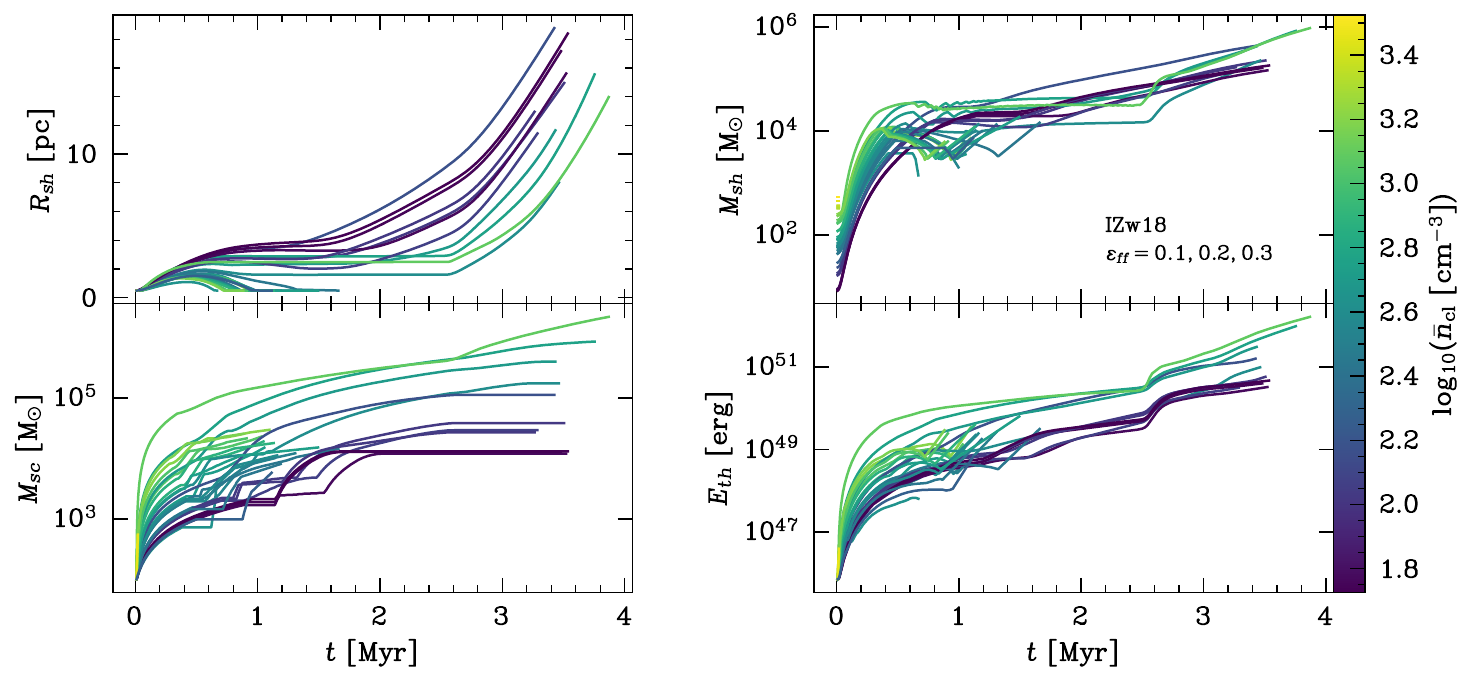}
\caption{Time evolution of the shell radius (top left), shell mass (top right), stellar mass (bottom left) and thermal energy (bottom right) for the low metallicity (IZw18) models, that were performed for three different values of $\epsilon_{ff}$ (the first three models listed in Table \ref{tab:sim_params}). The color map shows the average initial gas number density of the clouds, $\bar{n}_{\rm{cl}}=3 M_{\rm{gas,0}}/(4 \pi \mu m_{\rm H} R_{\rm{cl,0}}^3)$.}
\label{IZw18_all}
\end{figure*}

Note, however, that neither the shocked-wind temperature nor the density is uniform in the hot bubble. Hence, the energy loss rates must be calculated numerically and we do so by using our own numerical method that self-consistently estimates the radiative cooling rate of the bubble interior (see also \citealt{2019MNRAS.483.2547R}). The approach follows the general framework of \citet{Weaver1977}, in which mass evaporation from the swept-up shell feeds cooler gas into the hot interior, but we extend it to include time-dependent stellar input. At each timestep, we solve the coupled equations for the gas velocity and temperature in the region between $R_{\rm{rs}}$ and $R_{\rm{sh}}$, including the effects of heat conduction and gas cooling. The mass evaporation rate from the shell, $\dot{M}_{\mathrm{ev}}$, is obtained iteratively so that the integrated solution satisfies the required boundary conditions at both interfaces. Once convergence is reached, the term $n^{2}\Lambda(T)$ in the equations (see Equation \ref{mass_eva_eq2}) provides the total cooling rate $Q$ that enters Equation \ref{energy_eq}.

The implemented method naturally captures the transition from the early adiabatic phase to the radiative regime, without the need for ad hoc assumptions. Its performance was validated against 1D hydrodynamical simulations performed with the FLASH code \citep{fryxell_flash_2000}, adopting time-dependent stellar inputs from the BoOST evolutionary tracks. Both calculations yield consistent thermal energy evolution (see Fig. \ref{flash_compa} and \ref{flash_compa2}). A full description of the governing equations, boundary conditions, and numerical procedure is given in Appendix~\ref{app_evapora}.
\begin{table}
\caption{Simulation sets explored in this work.}
\label{tab:sim_params}
\centering
\begin{tabular}{l r}
\hline
Track / Metallicity & $\epsilon_{\rm ff}$ \\
\hline
IZw18 ($Z=0.00021$) & 0.1 \\
                    & 0.2 \\
                    & 0.3 \\
\hline
dwarfA ($Z=0.00105$) & 0.01 \\
                     & 0.03 \\
\hline
MW ($Z=0.0088$) & 0.01 \\
                & 0.03 \\
\hline
\end{tabular}
\tablefoot{The metallicity tracks are taken from \citet{BoOST2022}. The corresponding star formation efficiencies per free-fall time, $\epsilon_{\rm ff}$, are listed for each case. All simulations span the ranges $5.2 \leq \log[M_{\rm gas,0} (M_\odot)] \leq 6.5$ and $14 \leq R_{\rm cl,0} \,(\rm pc) \leq 31$.}
\end{table}

\subsection{Initial conditions and the selected parameter space}\label{initial conditions}
An initial stellar population (with mass $M_{\rm{sc}}(t_{\mathrm{ini}})=100$ M$_{\odot}$) is assumed to be formed at the initial time $t_\mathrm{ini}$. The initial gas mass ($M_{\rm{gas,0}}$), radius ($R_{\rm{cl,0}}$) and metallicity are the input parameters. It is assumed that $R_{\rm{c,0}}=0.2 R_{\rm{cl,0}}$ in all our calculations. The core gas density $\rho_{c}$ is calculated by integrating the density distribution. The shell is assumed to be located\footnote{See section \ref{sec_diff_r0} for a discussion on why we selected these values as initial conditions.} at $r_0=0.5$ pc at $t_{\mathrm{ini}}$. The initial shell mass $M_{\rm{sh}}(t_\mathrm{ini})$ is the swept-up gas cloud up to $r_{0}$. The initial shell velocity is assumed to be $U=0$ and we set the initial time to $t_{\mathrm{ini}}=10^4$ yr, in order to use the adiabatic self-similar solution by \cite{Weaver1977} to set the initial value of the thermal energy. 

For each choice of metallicity and $\epsilon_{\rm{ff}}$ given in Table \ref{tab:sim_params} we perform approximately $10^{3}$ simulations for varying values of $M_{\rm{gas,0}}$ and $R_{\rm{cl},0}$ within $5.2 \leq \log[M_{\rm{gas,0}} (M_{\odot})] \leq 6.5$ and $14 \leq R_{\rm{cl,0}} (\rm{pc}) \leq 31$, respectively.
\begin{figure}[htb]
\centering
	\includegraphics[width=\columnwidth]{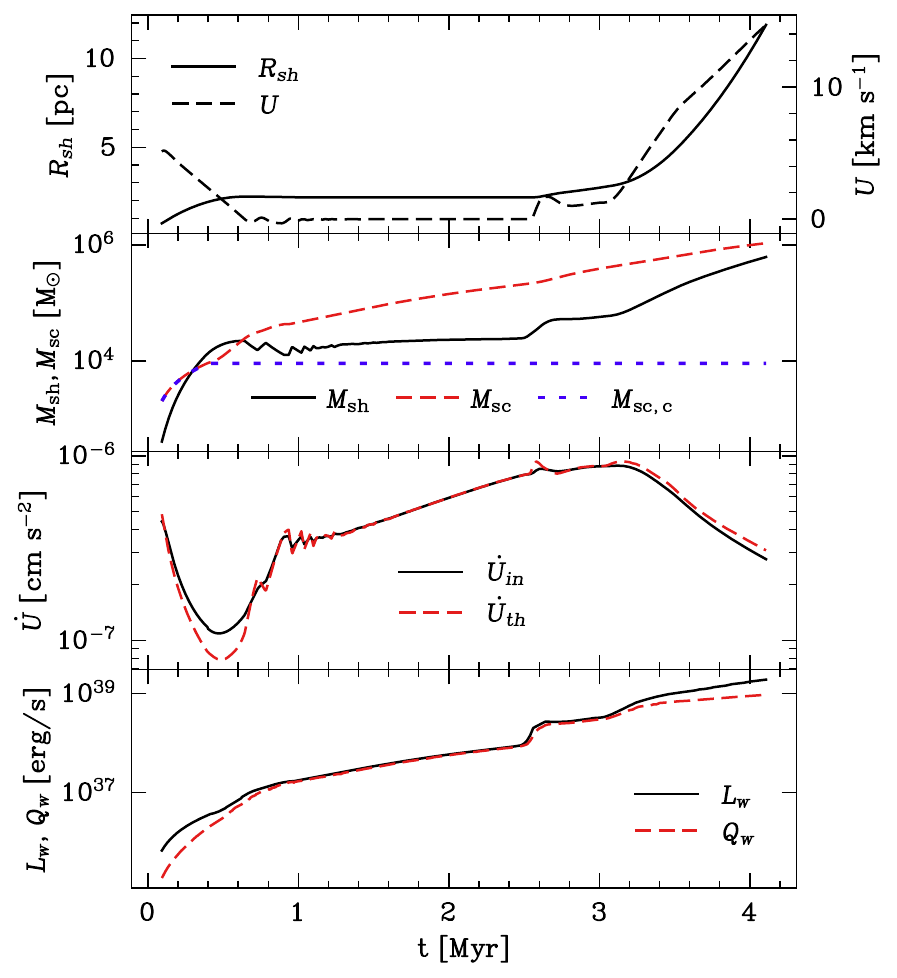}
    \caption{Model evolution for a case with $\log[M_{\rm{gas,0}} (M_{\odot})]=6.3$, $R_{\rm{cl}}=25.7$ pc, $\epsilon_{\rm{ff}}=0.3$ and for the IZw18 metallicity. First plot from top to bottom: shell radius (left y-axis), shell velocity (right y-axis). Second: shell mass ($M_{\mathrm{sh}}$), stellar mass formed from core star formation ($M_{\mathrm{sc,c}}$), and total stellar mass ($M_{\mathrm{sc}}$). Third: the outward force produced by the thermal energy ($\dot{U}_{\rm{th}}$, dashed) and the total inward forces ($\dot{U}_{\rm{in}}$, solid). Fourth, bottom right panel: Mechanical power (solid) produced by the stellar feedback and the cooling rate (dashed).}
\label{single_SSF}
\end{figure}

\section{Results} \label{sec_Results}
 \begin{figure}[htb]
\centering
	\includegraphics[width=\columnwidth]{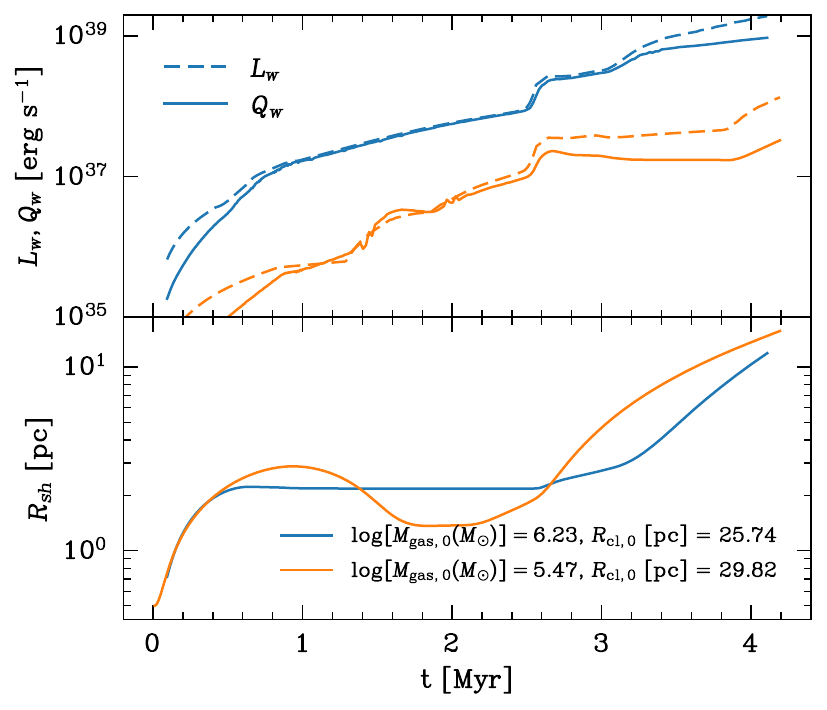}
\caption{Same as Fig. \ref{single_SSF}, but for two cases with different cloud gas masses and radii, as shown in the inset ($\epsilon_{\mathrm{ff}}=0.3$ in both cases).  Upper panel: time evolution of $L_{\rm{w}}$ (dashed lines) and $Q_{\rm{w}}$ (solid lines). Bottom panel: Evolution of the shell radius with time for the same models, using the same color as in the top panel. }
\label{figa_compare_stable}
\end{figure}
Fig. \ref{IZw18_all} presents the bubble evolution for a subset of the models for the lowest-metallicity case (see Appendix \ref{plots_all_results} for the full results for all cases listed in Table \ref{tab:sim_params}), and the three values of $\epsilon_{\rm{ff}}$ considered for this case (the first three entries in Table \ref{tab:sim_params}). The shell radius (top left), shell mass (top right), stellar mass (bottom left) and thermal energy (bottom right) are shown for each one of the models, and the color bar presents the average number density of the initial cloud, $\bar{n}_{\rm{cl}}=3 M_{\rm{gas,0}}/(4 \pi R_{\rm{cl,0}}^3)$. The calculations are always stopped either at shell collapse or when the shell has reached the outer boundary of the collapsing cloud. Note that the star formation time-scale for this case ranges from 0.5 to $\sim 5$ Myr, with final stellar masses ranging from $10^3$ to $10^6$ M$\odot$. The most massive clusters form in the densest clouds (bottom-left panel). This results from the higher core (Equation \ref{core_SFR}) and shell (Equation \ref{cond_fragmentation}) star formation rates (SFRs) that dense environments produce. The higher stellar masses ($M_{sc}$) in these cases also provide a stronger supply of mechanical power, leading to correspondingly higher thermal energies (bottom-right panel of Fig. \ref{IZw18_all}).
\begin{figure*}[htb]
\centering
	\includegraphics[width=1.6\columnwidth]{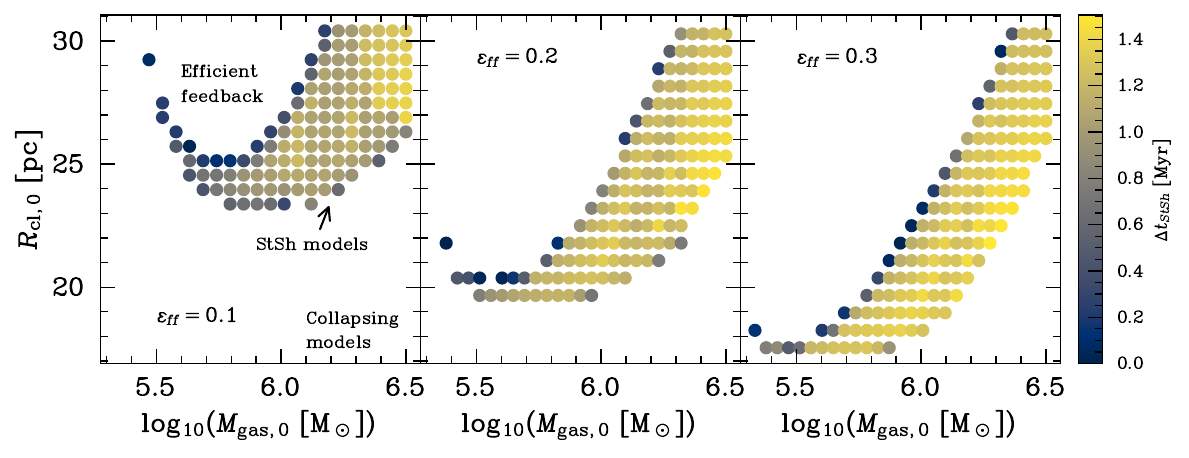}
\caption{Standing shell models for the IZw18 metallicity, plotted in the parameter space of the initial cloud radius and mass, with the subplots showing the different values of $\epsilon_{\rm{ff}}$ adopted for the calculations. Each model is color-coded with the duration of the StSh phase, $\Delta t_{\rm{StSh}}$ (see the color bar).}
\label{par_space_IZw18}
\end{figure*}

As the evolution of $R_{\rm{sh}}$ shows (top left panel in Fig. \ref{IZw18_all}), there are three different types of solutions for this metallicity case. First, models in which the feedback is unable to lift the shell, leading to its rapid infall on short timescales ($\sim 0.5$–$2$ Myr). This regime occurs in the densest clouds of our parameter space, where the feedback cannot overcome the high ram pressure of the collapsing cloud and the enhanced radiative cooling within the more confined bubble. As a result, the shell ultimately collapses. We do not analyze this regime in detail here, as it will be the focus of a forthcoming study. Nevertheless, it represents a particularly interesting case, since such clouds are expected to have multiple episodes of star formation. This behavior may be relevant, for example, to the formation of globular clusters, where multiple, chemically distinct stellar populations are commonly observed.

The second type of solution corresponds to models in which stellar feedback efficiently clears the surrounding gas, thereby quenching further star formation. In these cases, star formation stops rapidly, and the integrated star formation efficiency of the cloud remains low. This solution is characteristic of the lowest-density clouds in the models shown in Fig. \ref{IZw18_all}, but it also becomes the dominant outcome in higher-metallicity environments, where feedback is more effective (see the discussion in Section \ref{MW_metallicity_models}).

Finally, the third type of solution corresponds to models in which the expanding shells stall and reach a quasi-stationary equilibrium position. These standing shell (StSh) models follow the framework described by \citet{2003A&A...411..397T} and will be discussed in more detail in the following sections.

\section{The Star Formation Factory}\label{SFF_section}
Fig. \ref{single_SSF} presents the evolution of a single model of Fig. \ref{IZw18_all}, with input parameters $\log[M_{\rm{gas,0}} (M_{\odot})]=6.3$, $R_{\rm{cl}}=25.7$ pc, $\epsilon_{\rm{ff}}=0.3$ and IZw18 metallicity. From top to bottom, the first panel shows the shell radius and velocity, the second panel shows the shell mass ($M_{\mathrm{sh}}$), stellar mass formed from core star formation ($M_{\mathrm{sc,c}}$), and total stellar mass ($M_{\mathrm{sc}}$). The third panel presents the outward acceleration produced by the thermal energy ($\dot{U}_{\rm{th}}$) and the sum of all the inward acceleration terms, $\dot{U}_{\rm{in}}$. These are the positive and negative terms appearing on the right-hand side of Equation \ref{mod1}, respectively. Finally, the fourth (bottom) panel presents the evolving mechanical power and cooling rate, as indicated by the legend.

The shell initially expands, but soon stalls (see the first panel) due to the combined effects of the cloud ram pressure, gravity, and the increasing cooling rate (third panel). Although these inward forces would eventually cause the shell to collapse, around 0.4 Myr (When the shell expansion velocity drops to approximately
$0~\mathrm{km\,s^{-1}}$), it becomes gravitationally unstable. In this limit ($U \rightarrow 0$), equation~(\ref{star_formation_recipe}) reduces to $\omega_p \rightarrow \pi G \Sigma_{\rm sh} / c_{\rm s}$, implying rapid growth of gravitational perturbations and subsequent fragmentation into stars (see Section~\ref{add_star_formation}). This onset is marked by a slight decrease in shell mass and a corresponding increase in stellar mass, as shown in the second panel. In addition, the onset of star formation in the shell is also noticeable as the time when the total stellar mass ($M_{\mathrm{sc}}$) starts differing from the mass formed from core star formation ($M_{\mathrm{sc,c}}$), as shown in the second panel. Furthermore, note that by the end of the calculations, $M_{\mathrm{sc}}$ is nearly two orders of magnitude larger than $M_{\mathrm{sc,c}}$, thus showing that most of the star formation in our models occurs at the fragmenting shell.

The newly formed stars inject energy back into the bubble, causing a temporary re-expansion of the shell and inducing oscillations before it settles into an equilibrium configuration. The shell maintains a stationary position for a non-negligible period of time ($\sim 1.5$ Myr for this case). Interestingly, as also noted by \cite{2003A&A...411..397T}, during this phase the shell mass (solid line in the second panel) stays nearly constant while the stellar mass continues to rise. This indicates that the wind-blown bubble is continuously processing the collapsing cloud through its outer shell, efficiently converting the accreted gas into stars. Such a process leads to a highly efficient and compact mode of star formation, which \cite{2003A&A...411..397T} referred to as the Star Formation Factory, characterized by high star formation efficiency and long-term cluster survivability (see the discussion in Section \ref{results_trends}).

There are differences among the various StSh solutions. To illustrate this, Fig. \ref{figa_compare_stable} shows two additional models from Fig. \ref{IZw18_all}, with different initial cloud properties. The upper panel presents the mechanical power ($L_{\rm{w}}$, dashed) and the cooling rate ($Q_{\rm{w}}$, solid) while the bottom panel shows $R_{\rm{sh}}$ (with the same colors as in the upper panel). Note that the duration of the StSh phase differs among these models. In order to understand the differences, we performed a stability analysis of the system of equations (section \ref{sec_setup}) governing the bubble evolution (see Appendix \ref{standing_math}) and we demonstrate that a StSh solution is stable as long as:
\begin{equation}\label{cond_SS}
     \frac{1}{\tau} \int_{t_0}^{t_0+\tau} [L_{\rm{w}}(t')-Q_{\rm{w}}(t') ]\, dt' \to 0
\end{equation}
where $t_0$ is the onset of the equilibrium state and $\tau$ is the non-negligible duration of this phase. In other words, the standing solution is stable if the cooling rate in the hot bubble approaches the mechanical power. We verified numerically that, in the simulations, the average cooling rate approaches the mechanical power during these StSh phases. Note how the times when $L_{\rm{w}}$ decouples from $Q_{\rm{w}}$ in the top panel of Fig. \ref{figa_compare_stable} correspond to the instants when the shell starts expanding again (bottom panel). In physical terms, once the shell enters a StSh phase, it remains in that state until it can gain additional energy.

The reason why one of the models shown in Fig. \ref{figa_compare_stable} satisfies the condition given by Equation \ref{cond_SS} for a longer period than the other, lies in the cloud gas density. Indeed, in low density clouds the shell expands faster leading to a more rarefied bubble interior and thus to lower values of $Q_{\rm{w}}$. On the contrary, the higher cooling rate induced by the more compact structure, can keep up with the increasing values of $L_{\rm{w}}$ for a longer time-scale. Nevertheless, in all cases of the IZw18 metallicity, the StSh phase lasts only up to the onset of supernovae ($\sim 2.5$ Myr), since after that time all models explode given the power boost provided by the supernovae. 

It is also important to understand whether the existence of standing-shell (StSh) solutions is an artifact of our adopted star formation prescription, particularly the implementation of star formation within the shell. As discussed in Appendix \ref{star_formation_mode}, this evolutionary feature occurs for different star formation prescriptions. In particular, we demonstrate that StSh solutions appear even in models that assume purely continuous star formation in the core, without any contribution from shell star formation. Instead, we argue that this type of solution can arise commonly in star-forming collapsing gas clouds. 

\begin{figure}[htb]
\centering
	\includegraphics[width=\columnwidth]{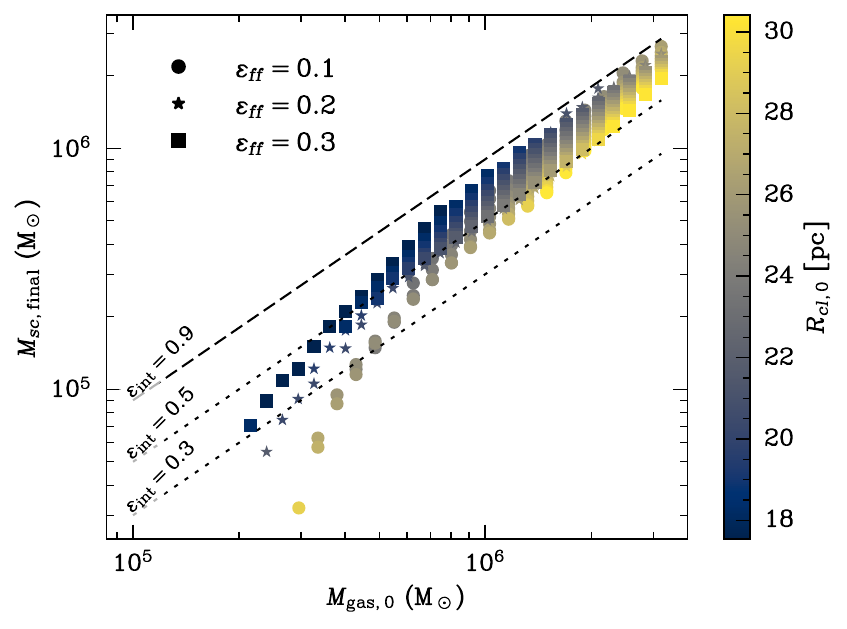}
\caption{The final stellar mass ($M_{\rm{sc,final}}$) obtained for IZw18 StSh models, as a function of the original cloud mass, $M_{\rm{gas,0}}$. We plot with different symbols the three values of $\epsilon_{\rm{ff}}$ for these calculations (as indicated by the inset legend). Each model color shows the initial cloud radius, $R_{cl,0}$, as shown by the color bar. Finally, the dashed lines present three different values (as shown in the figure) of the integrated star formation efficiency, i.e., $\epsilon_{\mathrm{int}}=M_{\rm{sc,final}}/M_{\rm{gas,0}}$. }
\label{eglobal_IZw18}
\end{figure}
\subsection{Global properties of Standing Shell models}\label{results_trends}
Fig. \ref{par_space_IZw18} presents all the StSh models found for the IZw18 case in the parameter space of the initial cloud properties (mass and radius), with the subplots showing the different values of $\epsilon_{\rm{ff}}$ adopted for the calculations. Each model is colored according to the duration of the StSh phase, $\Delta t_{\rm{StSh}}$ (shown in the color bar). The StSh models define an area of the parameter space, with collapsing models located below this area and continuously expanding shells (i.e., efficient feedback) located in the upper part of the parameter space (as shown by the inset labels on the left panel). Note that the StSh models shift toward the lower-right region of the plot (denser clouds) as $\epsilon_{\rm{ff}}$ increases. This is due to the fact that increasing the value of the star formation per free-fall time implies a higher stellar feedback from early on in the evolution of the bubbles, thus increasing the effectiveness of feedback on clearing the remaining gas cloud. Note that in the collapsing region of the parameter space,  more than one episode of star formation might be expected. For the $\epsilon_{\rm{ff}} = 0.1$ case, close to the observed values and to other numerical calculations \citep[e.g.][]{2011MNRAS.415.3439D,2024OJAp....7E.114P}, a large fraction of the parameter space is either StSh or collapsing, thus suggesting that in low metallicity collapsing clouds, star formation might be complex given the delayed effects of stellar feedback. Furthermore, the StSh stage time-scale is longer for denser models, in agreement with the discussion in Section \ref{SFF_section}.

Fig. \ref{eglobal_IZw18b} in the Appendix \ref{plots_all_results} shows that in the StSh models, star formation in the shell is the main contributor to the growth of stellar mass (for approximately two orders of magnitude compared to core star formation). Fig. \ref{eglobal_IZw18} shows the final, total stellar mass ($M_{\rm{sc,final}}$) obtained for the IZw18 StSh models, as a function of the original cloud mass, $M_{\rm{gas,0}}$, for the three values of $\epsilon_{\rm{ff}}$ considered for this metallicity (as indicated by the inset legend). Each model color shows the initial cloud radius, $R_{\rm{cl,0}}$ as shown by the color bar. Finally, the dashed lines indicate three different values (as shown in the figure) of the integrated star formation efficiency, $\epsilon_{\mathrm{int}}=M_{\rm{sc,final}}/M_{\rm{gas,0}}$, i.e., the total fraction of the original cloud that was transformed into stars by the end of the calculation. Note that, with the exception of low-mass models ($M_{\rm{gas,0}} \lesssim 4 \times 10^5$ M$_{\odot}$), all standing shell models produce similar final stellar masses, with the $\epsilon_{\mathrm{int}}$ in the range 0.5-0.9 for models with $M_{\rm{gas,0}} \gtrsim 10^6$ M$_{\odot}$. Thus, the Star Formation Factory indeed manages to produce stars very efficiently. 
\subsection{On the impact of cloud metallicity}\label{metal_impact}
The complete results for the dwarfA and MW metallicity cases (see Table \ref{tab:sim_params}) are presented in Appendix \ref{plots_all_results}. In this section, we first analyze the most important features of each metallicity case, followed by a discussion of the physical mechanisms underlying the differences observed between metallicities.

\subsubsection{MW metallicity}\label{MW_metallicity_models}
We did not find any StSh model for the MW metallicity (as shown in Fig. \ref{ap_model1}) for any of the two cases of the star formation efficiency per free-fall time that were considered for these calculations ($\epsilon_{\rm{ff}}=0.01, 0.03$). Star formation in these models occurs mostly between $1-2$ Myr and the thermal energies lie in the range $E_{\rm{th}}\sim 10^{51}-10^{53}$ erg. Stellar feedback is enough to produce continuously expanding bubbles, and unrealistically low values of $\epsilon_{\rm{ff}}$ would have to be considered to find StSh models for this case. The final, integrated star-formation efficiencies are very small, with an average of $\bar{\epsilon_{\mathrm{int}}} \sim 0.2$, but this is because we are considering massive and compact clouds. The average ranges between $\bar{\epsilon_{\mathrm{int}}} \sim 0.016-0.06$ for models with $M_{\rm{gas,0}} \lesssim 6 \times 10^{5}$ M$_{\odot}$, which is very close to the observed values in the MW and nearby local, star-forming galaxies.
\begin{figure}[htb]
\centering
	\includegraphics[width=\columnwidth]{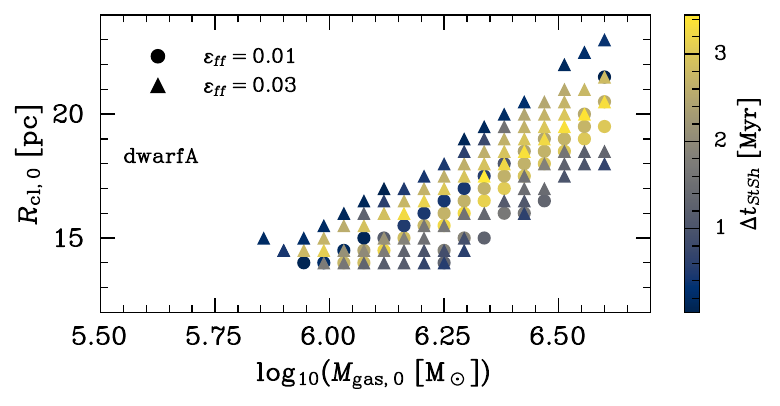}
        \caption{Same as Fig. \ref{par_space_IZw18} but for the dwarfA metallicity case and where both values of $\epsilon_{\rm{ff}}$ are plotted together (see the figure legend).}
\label{param_space_dwarfA}
\end{figure}
\subsubsection{dwarfA metallicity}
Fig. \ref{ap_model2} presents the calculations for the dwarfA metallicity. There are StSh models for this case. Star formation lasts up to 4 Myr, which is the longest time-scale for the three metallicities considered here, and also note that some of the StSh models collapse after 3-4 Myr of evolution. The stronger feedback in the dwarfA metallicity models results in StSh solutions being reached only for denser clouds compared to those in the IZw18 calculations. Figure \ref{param_space_dwarfA} shows the same figure as Figure \ref{par_space_IZw18}, but for the dwarfA metallicity and for both values of $\epsilon_{\rm{ff}}$ (see also Table \ref{tab:sim_params}). The dwarfA StSh models occupy roughly the same region of parameter space as the IZw18 StSh models, particularly those in the right panel of Figure \ref{par_space_IZw18} with $\epsilon_{\rm{ff}}=0.3$. However, the IZw18 models were computed using significantly larger $\epsilon_{\rm{ff}}$ values. In contrast, the dwarfA simulations adopt $\epsilon_{\rm{ff}}=0.01$–$0.03$; if higher efficiencies similar to those of IZw18 were used, the dwarfA StSh models would shift toward the lower-right region of the diagram, moving outside the range of the parameter space explored here. Finally, although with a larger scatter ($\epsilon_{\mathrm{int}} \sim 0.3-0.9$) than in Fig. \ref{eglobal_IZw18}, the final, integrated star formation efficiencies for dwarfA StSh are also large, with most models having $\epsilon_{\mathrm{int}} \sim 0.5-0.9$.
\begin{figure}[htb]
\centering
	\includegraphics[width=0.9\columnwidth]{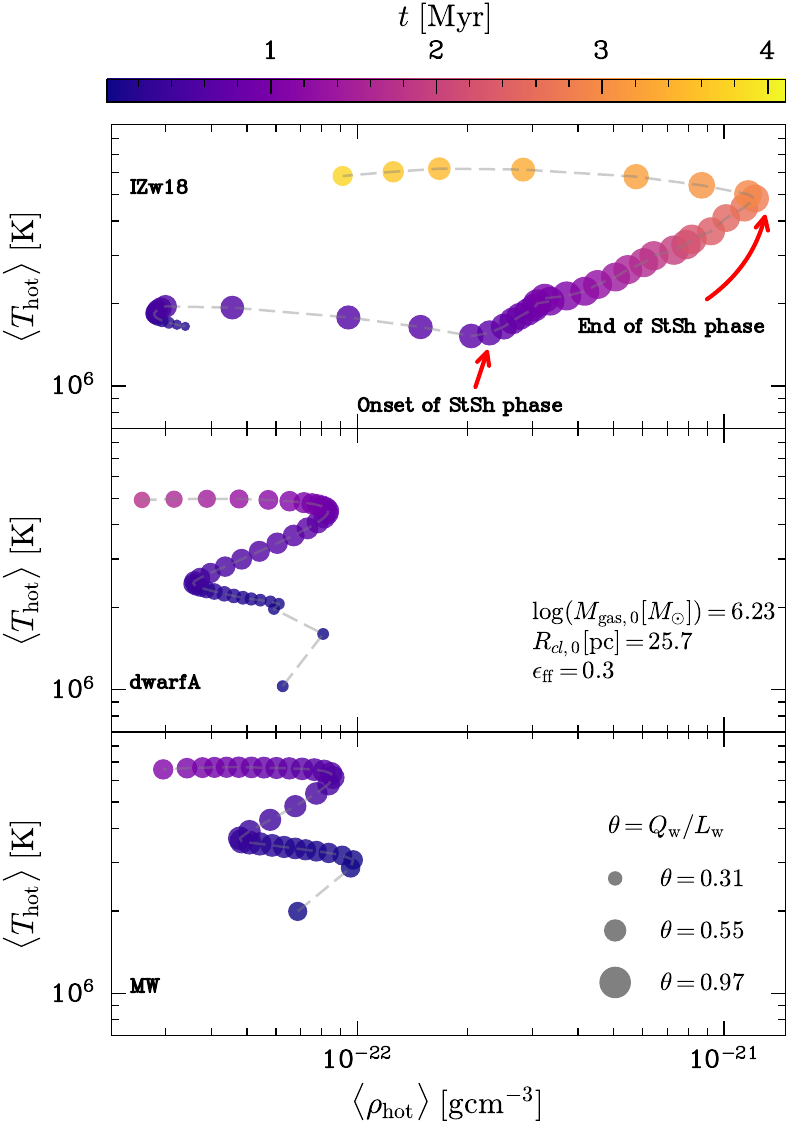}
        \caption{Phase space diagram, showing the emission-weighted density ($\langle \rho_{\mathrm{hot}} \rangle$) and temperature ($\langle T_{\mathrm{hot}} \rangle$) in the hot bubble. The size of the markers represents the feedback efficiency, defined here as $\eta_{\rm{feed}}=L_{\rm{w}}/Q_{\rm{w}}$, and the colorbar shows the time. The upper, middle and bottom panel present IZW18, dwarfA and MW metallicity simulations of the same cloud with initial properties: $\log (M_{\rm{gas,0}} [M_{\odot}])=6.23$, $R_{\rm{cl,0}} [\rm{pc}]=25.7$, and $\epsilon_{\rm{ff}}=0.3$. See the text for a full discussion on how these quantities were estimated.}
\label{phase_space}
\end{figure}

\subsubsection{The physics behind the metallicity differences}
The results discussed in the previous sections show, perhaps counterintuitively, that within massive, collapsing gas clouds, lower metallicity leads to longer star formation timescales and higher star formation efficiencies, even though cooling rates increase with metallicity. Indeed, one might expect that reduced cooling at low metallicity would enhance the impact of stellar feedback and therefore yield lower values of $\epsilon_{\mathrm{int}}$ in such environments. 

In order to understand our results, Fig. \ref{phase_space} shows a phase-space diagram of three cases. Each panel presents a model calculated with each one of the metallicities considered for this work (see the figure legends), and the remaining initial cloud properties are the same for the three panels: $\log (M_{\rm{gas,0}} [M_{\odot}])=6.23$, $R_{\rm{cl,0}} [\rm{ pc}]=25.7$, and $\epsilon_{\rm{ff}}=0.3$. For this particular cloud, a StSh phase is observed only in the IZw18 model, with onset and ending times shown in the figure (upper panel). The figure shows the emission-weighted average gas density and temperature ($\langle T_{\mathrm{hot}} \rangle$) in the hot bubble:
\begin{equation}\label{avg_T}
    \langle T_{\mathrm{hot}} \rangle= \frac{\int  Q_{\rm{b}} T \rm{d} V}{\int Q_{\rm{b}} \rm{d}V},
\end{equation}
\begin{equation}\label{avg_rho}
    \langle \rho_{\mathrm{hot}} \rangle= \frac{\int  Q_{\rm{b}} \rho \rm{d} V}{\int Q_{\rm{b}} \rm{d}V},
\end{equation}
where $Q_{\rm{b}}$ is the cooling rate as a function of radius. Note, for instance, that for a given time, the total cooling rate in the bubble that enters Equation \ref{energy_eq} is $Q_{\rm{w}} = \int Q_{\rm{b}} \rm{d}V$. The expressions \ref{avg_T} and \ref{avg_rho} are calculated on-the-fly during the calculations for each time. In addition, before plotting the results we applied a moving average with a window size of 10 data points. The marker sizes in the figure indicate the value of the feedback efficiency, defined as \citep[e.g.,][]{2019MNRAS.490.1961E, 2025ApJ...989...42L}:
\begin{equation}\label{feed_effi}
    \theta = Q_{\rm{w}} / L_{\rm{w}},
\end{equation}
as shown in the bottom panel legend. 

In the low metallicity calculation (upper panel), the bubble starts to cool, and around $\sim 1$ Myr, the increasing star formation rate due to the shell fragmentation, injects more energy and mass into the bubble, thus driving the increase of $\langle T_{\mathrm{hot}} \rangle$. Although mass evaporation from the cold shell into the hot bubble contributes to an increase in the mean inner density, $\langle \rho_{\mathrm{hot}} \rangle$, the reduced cooling in such a low-metallicity environment and the growing stellar mass, hinder the drop of the inner temperature. The marker sizes during the StSh phase indicate that $L_{\mathrm{w}} \sim Q_{\mathrm{w}}$ at this stage (as discussed in the previous sections). Subsequently, once supernovae begin to explode, $L_{\mathrm{w}}$ exceeds $Q_{\mathrm{w}}$, disrupting the previously established balance and driving a rapid, final expansion of the shell. This expansion results in a decrease of the inner density, evident in the figure as the final motion toward the left.

On the other hand, note that for larger metallicities (middle and bottom panels in Fig. \ref{phase_space}), the feedback efficiency is always $\theta<1$, i.e., the energy injection rate is always larger than the cooling rate for these cases. Hence, even though $Q_{\rm{w}}$ reaches larger absolute values at higher metallicity, the corresponding increase in the energy input rate surpasses $Q_{\rm{w}}$ throughout the entire evolution, resulting in a continuously expanding bubble. In these cases, the density within the bubble remains low given the continuous bubble expansion, further sustaining the cycle of weak cooling.

\section{Discussion} \label{sec_Discussion}
\subsection{Model limitations}
The one-dimensional nature of the present study inevitably implies certain limitations. In particular, our calculations cannot capture 3D processes such as turbulence, Rayleigh–Taylor and Kelvin–Helmholtz instabilities, or the development of asymmetric structures, which are intrinsic to the dynamics of collapsing clouds. Indeed, observed molecular clouds exhibit complex morphologies, including filaments, clumps, and over-dense regions \citep[e.g.][]{2022ApJ...929L..18E,2025A&A...693A.175P,2025MNRAS.541.3880Z}. These structural features are fundamental in shaping star formation \citep[e.g.][]{2025arXiv250919487G} and in determining how stellar feedback couples to the surrounding gas \citep[e.g.][]{2020MNRAS.499..748D,2025MNRAS.540.1124L}. Consequently, the resulting star formation efficiency in 3D simulations may differ from that obtained in our simplified framework. 

Our setup does not include turbulent mixing in the contact discontinuity, which is important on the calculation of the cooling rate \citep[e.g.][]{2019MNRAS.490.1961E,2024ApJ...970...18L, 2025ApJ...989...42L,2025ApJ...989...43L}. However, the model setup presented in this work includes a detailed calculation of the cooling rate in the hot bubble, including at the interface between the hot gas and the shell, which is important to determine the dynamical evolution of feedback-driven bubbles. Our method self-consistently estimates the mass evaporation rate from the cold shell into the hot bubble, taking into account radiative losses. This has been shown to change by a factor of $\sim 3-30$, the evaporation rates obtained by \cite{Weaver1977} in their analytical approximation \citep{2019MNRAS.490.1961E}. In addition, the thermal evolution in our models have been shown to be in good agreement with the results of full hydro simulations. 

In this work, radiation pressure is not modeled explicitly. While it can influence star formation efficiency and potentially affect the resulting stellar initial mass function (IMF) \citep[e.g.][]{2024ApJ...975L..17M,2024MNRAS.530.2453C}, previous studies suggest that its impact is most pronounced at very early evolutionary stages ($\lesssim 1$ Myr) or under conditions where the cluster’s effective wind power is substantially reduced. In particular, \citet{2014ApJ...785..164M} found that radiation pressure primarily changes the shell structure under these circumstances and tends to dominate only in very dense environments. 

Consistent with our focus on mechanical feedback, we adopted a fixed shell sound speed of $c_s = 1$ km s$^{-1}$, appropriate for neutral gas. If the shell were significantly ionized, the higher sound speed would modify its fragmentation properties (see equations \ref{star_formation_recipe}-\ref{star_formation_recipe2}) and could delay or regulate shell star formation. However, predicting the net effect of varying values of $c_s$ would require a fully self-consistent modeling, as delayed fragmentation may simultaneously enhance the relative importance of gravity during the evolution. A more comprehensive model including radiation feedback would naturally account for spatial variations in the sound speed across ionized and neutral regions of the shell (which in many cases dominates the shell mass) \citep[e.g.][]{2014ApJ...785..164M, 2023MNRAS.521.5686K}. 

However, since our parameter space focuses on later stages and physical regimes where wind-driven feedback is expected to dominate, the omission of radiation pressure is not expected to qualitatively affect our main conclusions. Nonetheless, incorporating radiation pressure effects represents an important extension for future developments of the model.

It is worth noting that models exhibiting prolonged star formation (e.g. the standing shell models) and consequently higher efficiencies in our 1D framework may correspond, in three dimensions, to cases where stellar feedback is also relatively inefficient. This could arise from higher radiative cooling and the escape of feedback energy through low-density channels. The conditions under which our models have standing phases and thus high star formation efficiency, namely high densities and low metallicities, are also expected at high redshift, where massive galaxies are believed to have formed rapidly and with high star formation efficiencies \citep[e.g.][]{2009Natur.457..451D,2025MNRAS.544..160D, 2025MNRAS.544.3774S}. 

Despite these limitations, the presented 1D calculations incorporate many of the key physical ingredients governing cloud evolution and star formation. It follows the gravitational collapse of a gas cloud, includes time-dependent stellar winds and supernova feedback, and accounts for gravitational instabilities and subsequent star formation within the swept-up shell. Moreover, the model includes a detailed calculation of gas cooling within the hot bubble, including the contribution from mass evaporation from the cold shell into the hot interior. These ingredients together allowed us to explore the interplay between cloud collapse, feedback, and the efficiency of star formation for a large set of models covering two orders of magnitude in metallicity.  

\subsection{Comparison to other works}
Although a direct comparison with other studies is difficult due to differences in physical assumptions and numerical implementations, several works on massive and compact star cluster formation are particularly relevant. \citet{2023MNRAS.521.5686K} modeled star cluster formation in collapsing gas clouds using one-dimensional FLASH simulations that included gas self-gravity and time-dependent stellar feedback from winds and radiation. Star formation was implemented self-consistently, occurring only when specific physical criteria were met.

\begin{figure}[htb]
\centering
	\includegraphics[width=\columnwidth]{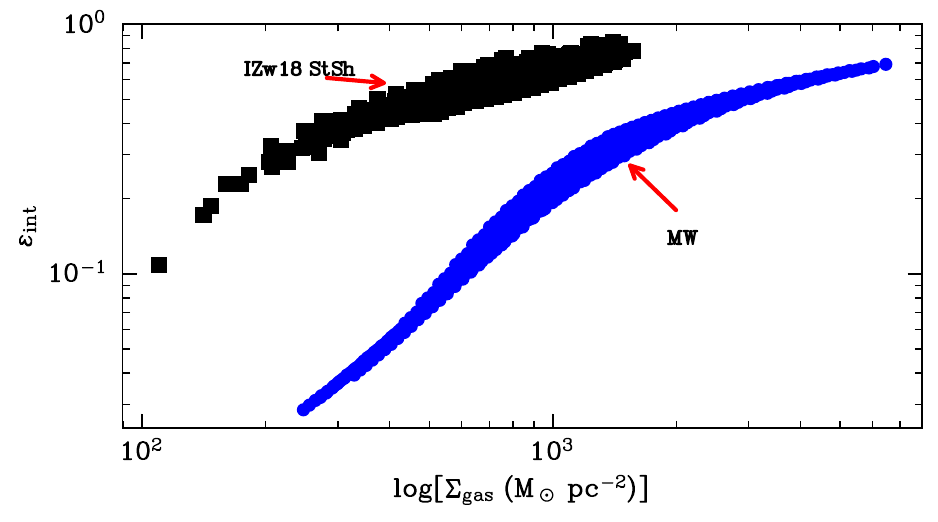}
        \caption{Integrated star formation efficiencies $\epsilon_{\mathrm{int}}$ as a function of the initial gas cloud surface density, $\Sigma_{\mathrm{gas}}$ for all the MW and the standing shell IZw18 models, as shown in the lower and upper parts of the plot, respectively.}
\label{discu_sigmas}
\end{figure}

While a detailed comparison is difficult, since their models did not include supernova feedback or shell fragmentation and we did not include radiation feedback, note that some of their simulations exhibit standing shell phases, suggesting that the StSH solution may be intrinsic to collapsing gas clouds. They also found that star formation efficiency correlates with cloud density. In our work, the efficiency correlates more strongly with surface gas density (see Fig. \ref{discu_sigmas}); however, \citet{2023MNRAS.521.5686K} also obtained high efficiencies for dense clouds, consistent with our results.

\cite{2024A&A...690A..94P} recently performed multi-physics simulations of star cluster formation by using the Torch framework. They studied three massive, turbulent, star-forming clouds ($M=10^4,10^5$, and $10^6$ M$\odot$) for a single cloud radius ($R=11.7$ pc). They assumed MW metallicity and found high integrated star formation efficiencies in their calculations ($\epsilon_{\mathrm{int}}=0.36, 0.65$ and $0.85$, respectively). They also argue that the integrated star formation efficiency depends mainly on the gas surface density. Fig. \ref{discu_sigmas} presents $\epsilon_{\mathrm{int}}$ as a function of $\Sigma_{\mathrm{gas}}$ for all of our MW and the standing shell IZw18 models, as indicated by the inset labels. Note that our models indeed show a strong correlation between these variables, in agreement with the findings by \cite{2024A&A...690A..94P}. However, our star formation efficiencies are a bit smaller than theirs, which is perhaps in better agreement with observations, such as in the case of MW clouds, where star formation efficiencies are $\epsilon_{\mathrm{int}}\lesssim 0.3$ \citep[e.g.][]{2003ARA&A..41...57L,2009ApJS..181..321E,2011ApJ...729...72P}. 

\cite{2018MNRAS.475.3511G} also found that the integrated star formation efficiency, $\epsilon_{\mathrm{int}}$, correlates most strongly with the gas surface density. Although their simulations included more physical processes than ours, our results are in good agreement, particularly for dense clouds. For example, at $\Sigma_{\mathrm{gas}} \sim 10^{3} \mathrm{M_{\odot}}$  pc$^{-2}$, our MW models produce $\epsilon_{\mathrm{int}}$ values in the range $0.19$–$0.25$, consistent with the $\epsilon_{\mathrm{int}} = 0.24$ obtained from the fit that \cite{2018MNRAS.475.3511G} made to their results. For our densest model, with $\Sigma_{\mathrm{gas}} \sim 6.5 \times 10^{3} \mathrm{M_{\odot}}$ pc$^{-2}$, we find $\epsilon_{\mathrm{int}} = 0.69$, compared to their slightly lower value of $\epsilon_{\mathrm{int}} = 0.58$. This might be due to the increased importance of radiation pressure at higher densities, which is not included in our modeling. In contrast, for lower-density clouds ($\Sigma_{\mathrm{gas}} \lesssim 7 \times 10^{2} \mathrm{M_{\odot}} $ pc$^{-2}$), our results align more closely with those of \cite{2011MNRAS.417..950H} than with \cite{2018MNRAS.475.3511G}. \cite{2018ApJ...859...68K} presented simulations including the effects of photo-ionization and radiation pressure. They also found a strong correlation between $\Sigma_{\mathrm{gas}}$ and $\epsilon_{\mathrm{int}}$. However, we obtained slightly smaller values of the integrated star formation efficiency. For instance, for $\Sigma_{\mathrm{gas}} \sim 10^{3} \mathrm{M_{\odot}}$  pc$^{-2}$ our obtained values for $\epsilon_{\mathrm{int}}$ ($0.19$–$0.25$) are smaller than the values obtained by \cite{2018ApJ...859...68K}  ($\epsilon_{\mathrm{int}} \sim 0.45-0.55$).

In addition to these studies, several other works have addressed this issue from different approaches \citep[e.g.][]{2013MNRAS.435.1701C, 2018ApJ...859...68K, 2020SSRv..216...50C, 2022ApJS..259...21K,2024Sci...384.1488F, 2025arXiv251007368D}. The results obtained from these works depend both on the simulations setup and on the included physics. Indeed, in addition to the prescription for feedback and the included mechanisms, the initial conditions for the cloud could also impact the obtained integrated results. For instance, clouds with a power-law density profile (as assumed in this work) might be easier to disrupt \citep{2010ApJ...709..191M} than uniform density clouds (as assumed by \citealt{2018MNRAS.475.3511G} and \citealt{2018ApJ...859...68K}) . Simulations including only radiation feedback seem to produce slightly larger $\epsilon_{\mathrm{int}}$ for dense clouds \citep[e.g.][]{2021MNRAS.506.5512F, 2023MNRAS.521.5160M}. However, all of these works coincide on establishing the surface density as the main parameter setting the integrated star formation efficiency in massive star-forming gas clouds \citep[e.g.][]{2025arXiv251007368D}.  

\subsection{Observations and low-metallicity environments}
There is evidence for the presence of massive star-forming regions in the local Universe. Examples include the starburst galaxies NGC 5253 \citep[e.g.,][]{2015Natur.519..331T}, NGC 253 \citep[e.g.,][]{2018ApJ...869..126L, 2020MNRAS.491.4573R}, and NGC 4945 \citep[e.g.,][]{2020ApJ...903...50E}, where star clusters with masses of $10^{4}$–$10^{6}$ M$_{\odot}$ are currently forming with high star formation efficiencies. Within our own Galaxy, massive cluster formation is also observed, for instance, in the well-studied W43 giant molecular cloud (GMC), which hosts a forming cluster of mass $\gtrsim 5 \times 10^{4}$ M${\odot}$ embedded in a gas reservoir of $\sim 10^{6}$ M$_{\odot}$ within a compact region of $\lesssim 60$ pc \citep[e.g.,][]{2013ApJ...779..121G, 2016ApJ...828...32L}.

Massive star formation is also prevalent in interacting galaxies, such as the Antennae Galaxies \citep[e.g.,][]{2014ApJ...795..156W}, where giant molecular clouds are found to coincide with sites of recent and intense star formation \citep[e.g.,][]{2012ApJ...750..136W}. A particularly and well studied case is the Firecracker cloud, a compact ($R \sim 20$ pc) and massive ($M \sim (1$–$9) \times 10^{6}$ M$_{\odot}$) star-forming region \citep{2019ApJ...874..120F} that appears to still be collapsing while actively forming new stars \citep{2024A&A...690A..94P}. Altogether, these examples demonstrate that very massive, compact, and efficient star formation sites can indeed arise in diverse galactic environments, consistent with the physical conditions and results explored in our models.

In low-metallicity environments, our results indicate that stellar feedback is substantially weaker, leading to higher star formation efficiencies. As illustrated in Fig. \ref{discu_sigmas}, the StSh models corresponding to our IZw18–like metallicity calculations exhibit significantly larger $\epsilon_{\mathrm{int}}$ values compared to the MW models at the same $\Sigma_{\mathrm{gas}}$. As discussed in Section \ref{results_trends}, denser models tend to develop collapsing shells, suggesting that multiple episodes of star formation may occur in these conditions.

At high redshift, low-metallicity environments are expected to host massive star-forming clouds \citep[e.g.,][]{2020ARA&A..58..157T,2026enap....4..500K}, where feedback suppression has also been predicted \citep[e.g.,][]{2023MNRAS.523.3201D,2024A&A...690A.108L}. Recent JWST observations have begun to reveal numerous compact and massive star-forming regions at high redshift \citep[e.g.,][]{2023MNRAS.526.2696H,2024MNRAS.531.4099R}, with evidence pointing toward reduced feedback efficiency in such environments \citep[e.g.,][]{2023ApJ...958..149J,2023ApJ...957...77P,2024ApJ...976..166P}. These observational trends are consistent with our findings that decreasing metallicity leads to longer evolutionary timescales and higher star formation efficiencies. Such behavior may be important to understand the physical conditions under which globular clusters formed in the early Universe.

\section{Summary and Conclusions}\label{sec_Conclusions}
Our aim in this work was to investigate the impact of stellar winds and supernova feedback on the star formation in collapsing gas clouds. To this end, we modeled the evolution of feedback-driven bubbles that are formed by the star-forming clusters within clouds of different masses, radii and metallicities. We have developed a one-dimensional code that performs calculations including important physical processes, such as the time-dependent feedback parameters (mass injections and mechanical power), star formation, gravity, cloud collapse, shell gravitational instabilities and triggered star formation, as well as radiative cooling in the hot bubble. The cooling term also accounts for the heat transfer between the cold and hot interfaces in the contact discontinuity. Our numerical setup was tested against hydrodynamical simulations performed with the FLASH code, showing good agreement while being much faster. Using this setup, we have performed an extensive exploration of the parameter space covering a broad range of the initial cloud masses, radii and metallicities. We list our main conclusions as:
\begin{enumerate}
      \item Star formation in high metallicity (MW-like) clouds occurs during short timescales ($\sim1-2$ Myr) and the final star formation efficiency, defined as the fraction of the original cloud transformed into stars, ranges between $0.016-0.06$ for models with masses $\lesssim 5 \times10^6$ M$\odot$ and up to 0.2 for more massive clouds. Hence, stellar feedback is efficient in clearing the collapsing gas cloud in a continuously expanding wind-blown bubble.
      \item For lower metallicities, we observed two additional regimes. Indeed, in addition to the efficient feedback regime (as for MW metallicities), we found for low metallicity clouds, that models can also collapse or enter a standing-shell phase. 
      \item The collapsing models are found for the densest clouds in our studied parameter space. This occurs because the insufficient feedback cannot withstand the several opposing processes, namely, gravity, ram pressure of the collapsing cloud and radiative cooling in the bubble interior. Such collapsing models, although here not studied in detail, may lead to several episodes of star formation and hence to star clusters hosting multiple generations of stars \citep{2018MNRAS.473L..11R}.  
      \item The third regime found in our calculations, is the standing shell solution. In this case, the shell evolves in such a way that it reaches pressure balance between the outward force created by the thermal energy and the inward forces. During this stage, the shell is gravitationally unstable, thus it fragments and forms stars that contribute to the feedback budget. Our models show that, if the shell reaches a standing position, the shell mass is approximately constant while the cloud is being accreted onto the shell. This means that the feedback-driven shell efficiently converts gas out of the collapsing cloud, thus leading to a massive, efficient, and compact mode of star formation. 
      \item In particular, standing solutions are commonly found for the lowest metallicities, with a significant portion of the parameter space displaying this solution. For the intermediate metallicities, the occurrence of this phase is only possible for the densest models for the adopted values of the star formation per free-fall time, those with initial cloud masses and radii in the range $\log[M_{gas,0} (M_{\odot})] \sim 6.0-6.5$, and $R_{cl,0} \sim 15-20$ pc, respectively.  Furthermore, we showed that this star formation mode leads to high final star formation efficiencies, with values in the range $0.5-0.9$. 
      \item We find that the decrease in mechanical power with metallicity has a stronger effect than the associated reduction in cooling efficiency, leading to more prolonged and efficient star formation in low-metallicity environments.
      
\end{enumerate}

Despite the simplifications inherent in our one-dimensional model, our results highlight the important role of metallicity in regulating the evolution and ultimate outcome of star formation in collapsing massive gas clouds.

\begin{acknowledgements}
The authors thank the anonymous referee for helpful comments and suggestions that improved the quality of this paper. This work was supported by the Czech Ministry of Education, Youth and Sports, through the INTER-EXCELLENCE II program, project LUC24023 (MSMT-14950/2024-5). We also acknowledge support from the institutional project RVO:67985815. S.J. acknowledges the support provided by the Czech Academy of Sciences, through its Programme to Support Prospective Human Resources – Postdoctoral Fellows (PPLZ), contract number L100032351. {\it Software:} a part of this work used FLASH v4.6, which was partly developed by the DOE NNSA-ASC OASCR
Flash Centre at the University of Rochester \citep{fryxell_flash_2000}. We also used NumPy \citet{Numpy}, Windcalc \citep{Wunschetal2017}, Matplotlib \citep{Matplotlib}, smplotlib \citep{jiaxuan_li_2023_8126529}, and SciPy \citep{SciPy}.
\end{acknowledgements}
\bibliographystyle{aa}
\bibliography{ref.bib}

%
%

\begin{appendix}
\renewcommand\thesection{A}
\section{The calculation of gas cooling in the hot wind bubble}\label{app_evapora}
The cooling rate in the hot wind bubble ($Q$ in Equation \ref{energy_eq}) is estimated following \cite{Weaver1977}. Indeed, for each time step (t), the radial velocity and temperature profiles within the hot bubble, specifically in the region between the reverse shock ($R_{\rm{rs}}$) and the outer shell radius ($R_{\rm{sh}}$), $R_{\rm{rs}} \leq x \leq R_{\rm{sh}}$,  are the solution to the equations:
\begin{equation}\label{mass_eva_eq1}
    \frac{1}{x^{2}}\frac{\partial}{\partial x} \left( x^2 v\right)-\left(v-\frac{\alpha x }{t}\right)\frac{1}{T}\frac{\partial T}{\partial x}=\frac{\beta+\delta}{t},
\end{equation}
\begin{align}
\frac{1}{P_{\rm{th}} x^2} \frac{\partial}{\partial x} 
\left( C T^{5/2} x^2 \frac{\partial T}{\partial x} \right)
& - \frac{5}{2} \left( v- \frac{\alpha x}{t} \right)
\frac{1}{T}\frac{\partial T}{\partial x} \notag \\
& - \frac{n^2 \Lambda}{P_{\rm{th}}} = \frac{\beta+2.5 \delta}{t},
\label{mass_eva_eq2}
\end{align}
where v, n and T are the flow velocity, number density and temperature, $C=6 \times 10^{-7}$ erg s$^{-1}$ cm$^{-1}$ K$^{-7/2}$ is the coefficient of thermal conductivity \citep{2019MNRAS.490.1961E} and:
\begin{equation}
    \alpha= \frac{\rm{d} \ln R_{\rm{sh}}}{\rm{d} \ln t}, 
\end{equation}

\begin{equation}
    \beta=-\frac{\rm{d} \ln P_{\rm{th}}}{\rm{d} \ln t}, 
\end{equation}
and
\begin{equation}
\delta=\frac{\rm{d} \ln T}{\rm{d} \ln t}.
\end{equation}

Here we describe the method developed to estimate the cooling rate in the hot bubble. Equations \ref{mass_eva_eq1}--\ref{mass_eva_eq2} are integrated numerically at every time step, to find the velocity and temperature profiles within the hot wind bubble, i.e., for $R_{\rm{rs}} \leq x \leq R_{\rm{sh}}$. The thermal pressure is assumed uniform in the interior and the integration starts from a position near the outer radius and is performed backwards up to the position of the reverse shock. The boundary conditions are given by:
\begin{equation}
\lim_{x \to R_{\rm{sh}}} T = 0
\end{equation}

\begin{equation}
\lim_{x \to R_{\rm{sh}}} v = U
\end{equation}

\begin{equation}\label{boundary3}
\lim_{r \to 0} v = 0
\end{equation} 
The mass evaporation from the shell into the bubble is related to the boundary conditions by:
\begin{equation}
    \frac{\rm{d}M_{\rm{b}}}{\rm{d}t} =\lim_{x \rightarrow R_{\rm{sh}}} 4 \pi x^{2} \frac{\mu P_{\rm{th}}}{k} \frac{\alpha R_{\rm{sh}} /t -v}{T}.
\end{equation}
However, the final solution must be found by iteration. Indeed, the method described by \cite{Weaver1977} is to iterate the calculation with varying values of $\rm{d} M_{\rm{b}} /\rm{d}t$, until the velocity found by numerical integration satisfies the third boundary condition (Equation \ref{boundary3}). Hence, for every time step, the iteration starts with the initial value of the mass evaporation: 
\begin{equation}
    \frac{\rm{d} M_{\rm{b}}}{\rm{d}t} \sim  \frac{7 \mu r^3}{t k_{\rm{B}}} \left( \frac{t C}{r^2} \right)^{2/7}P_{\rm{th}}^{5/7},
\end{equation}
which is the analytical adiabatic solution, and then this quantity is iterated until the boundary conditions are satisfied. For each value of $\rm{d}M_{\rm{b}} /\rm{d}t$, the initial conditions for the differential equations are:

\begin{equation}
    T=\left ( \frac{25}{4} \frac{k_{\rm{B}}}{\mu C} \frac{d M_{\rm{b}}/dt}{4 \pi R_{\rm{sh}}^2}\right)^{2/5} \left( R_{\rm{sh}} -x \right)^{2/5},
\end{equation}
\begin{equation}
    \frac{\partial T}{\partial x}=-\frac{2}{5}\frac{T}{R_{\rm{sh}}-x},
\end{equation}
\begin{equation}
v=\frac{\alpha r}{t}-\frac{\rm{d} M_{\rm{b}}/\rm{d}t }{4 \pi R_{\rm{sh}}^2}\frac{k_{\rm{B}} T}{\mu P_{\rm{th}}}.    
\end{equation}
To carry out the iteration and find the best value of $\rm{d}M_{\rm{b}} /\rm{d}t$ that satisfies all the boundary conditions, we have implemented a binary search algorithm. Furthermore, we have used this calculation to determine the cooling rate ($Q$) for the entire bubble region at each time step, since the term $n^{2}\Lambda$ is calculated when solving the equations (see Equation \ref{mass_eva_eq2}).

In order to test our code, we performed a calculation of the bubble evolution produced by a single massive star, which expands into a uniform medium with number density $n_0=100$ cm$^{-3}$. The time evolution of the star's mechanical power and mass injection rate are taken from the BoOST stellar tracks \citep{BoOST2022}. This calculation is compared to a simulation performed with the FLASH hydro code \citep{fryxell_flash_2000}. The hydrodynamic equations were solved using a modified version of the piecewise parabolic method \citep{colella_piecewise_1984}. We used the same cooling function by \citet{Schure2009} and the same prescription for thermal conduction as described above. The diffusion equation was solved using an implicit backward Euler scheme. We disabled radiative cooling for the initial \(50\,\mathrm{kyr}\) of the simulation to prevent unphysical cooling before the bubble's structure developed. We implemented the stellar wind by inserting mass and thermal energy into a 9-cell radius sphere at the center following the generalized Schuster profile \citep{ninkovic_generalised_1998}. The simulation was performed using a 1D, spherically symmetric grid of 16384 cells spanning \(45\,\mathrm{pc}\), with reflect boundary conditions at \(r = 0\) and diode boundary conditions at the outer edge. Despite the high resolution, the contact discontinuity was not fully resolved. Thermal conduction smooths out discontinuities in temperature into steep gradients. At the contact discontinuity, the temperature crosses the peak of the cooling function. Being time-step limited, we were not able to fully resolve this cooling feature. Consequently, there is moderate overcooling, resulting in a bubble with less thermal energy and mass but a larger radius than the correct result.

\begin{figure}[htb]
\centering
	\includegraphics[width=\columnwidth]{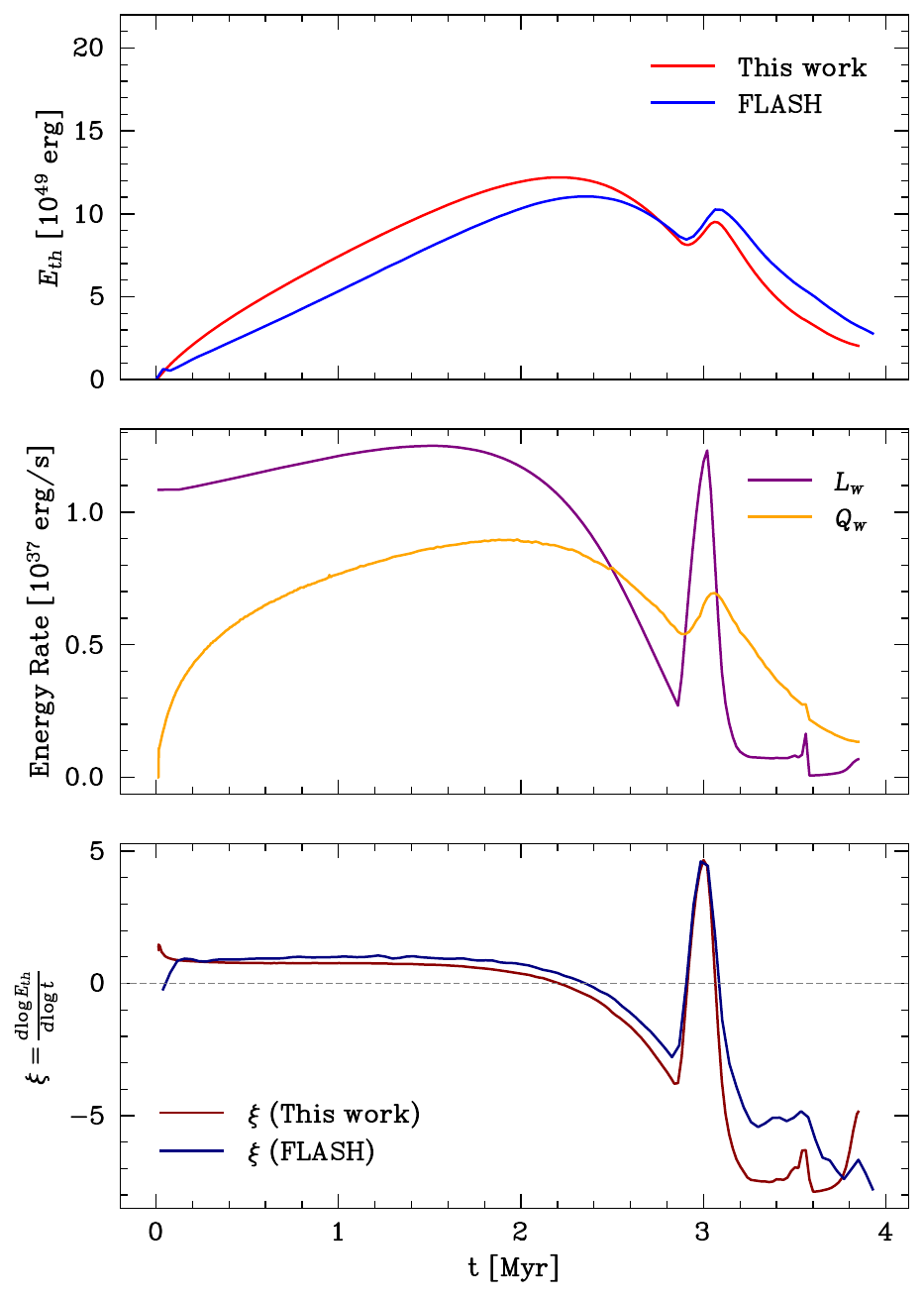}
\caption{Top panel: the thermal energy evolution of the wind-blown bubble driven by the feedback of a single massive star. The different curves present our results and those from a FLASH simulation performed with a refinement level 12 (see the legend). Middle panel: the star's mechanical power injection rate as a function of time as given by the BoOST stellar tracks \cite{BoOST2022}, and the cooling rate calculated by our approach. Bottom panel: a comparison between ours and the FLASH  results for the power-law index of the thermal energy evolution $\xi=\rm{d} \log E_{\rm{th}} / \rm{d} \log t$. See the text for a discussion.}
\label{flash_compa}
\end{figure}

Fig. \ref{flash_compa} presents the results. The top panel shows the time evolution of the thermal energy, both for our method and for the FLASH simulation (see the legend). This figure shows that our method produces an evolution of $E_{th}$ that is consistent with the values obtained from the full hydro simulation, the two results agree considerably well during the entire evolution of the model. The small differences in both calculations are probably due to their different initial conditions.  The middle row presents the star's mechanical energy injection rate as given by the BoOST stellar tracks \cite{BoOST2022}, and the cooling rate produced by our calculation. Note that, as expected, the bubble makes a transition from the early adiabatic stage to the radiative evolution, and only around $\sim 2$ Myr, the cooling rate is large enough to produce a drop in the thermal energy.  The bottom row shows a comparison from the power-law index of the thermal energy evolution, $\xi= \rm{d} \log E_{\rm{th}} / \rm{d} \log t$, for both calculations. This also confirms that both calculations lead to similar evolution for $E_{\rm{th}}$.

\begin{figure}[htb]
\centering
	\includegraphics[width=\columnwidth]{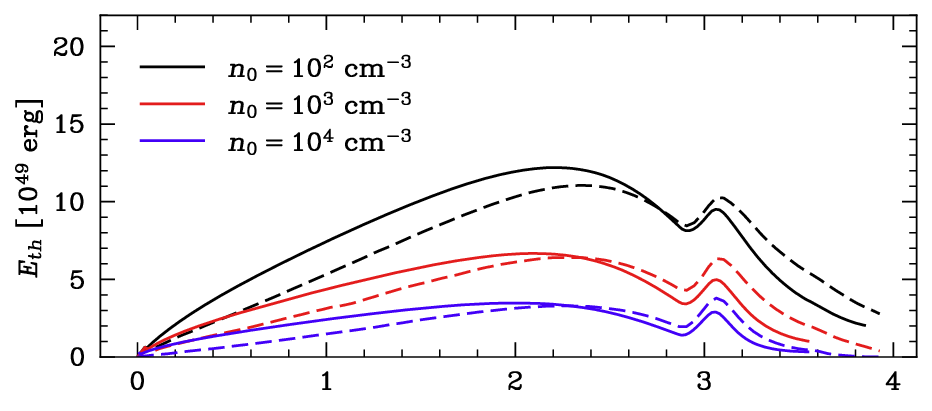}
\caption{The thermal energy as a function of time for FLASH (dashed) and this work (solid) calculations for different values of the ambient gas density (as shown in the legend). The remaining parameters and initial conditions from the calculations are the same as in Fig. \ref{flash_compa}.}
\label{flash_compa2}
\end{figure}

Fig. \ref{flash_compa2} presents the same as the top panel in Fig. \ref{flash_compa}, but for additional values of the ambient gas density (as indicated by the legend). This figure indicates that our numerical method manages to reproduce FLASH simulations even in high density media, which is important given the large range of densities of the star-forming gas clouds considered in this work. 

\renewcommand\thesection{B}
\section{The initial conditions for our calculations}\label{sec_diff_r0}
As stated in section \ref{sec_struct_shells}, our calculations assume that the outer shell of swept-up material has already cooled down and thus remains thin compared to the shell radius. In addition, our models assume the same initial position of the shell, $r_0=0.5$ pc, for all models. In reality, the radius and time at which a thin shell forms are expected to depend on the physical properties of the cloud \citep[e.g.,][]{1992ApJ...388...93K, 1992ApJ...388..103K, 2025ApJ...989...42L}. 

\begin{figure}[htb]
\centering
	\includegraphics[width=\columnwidth]{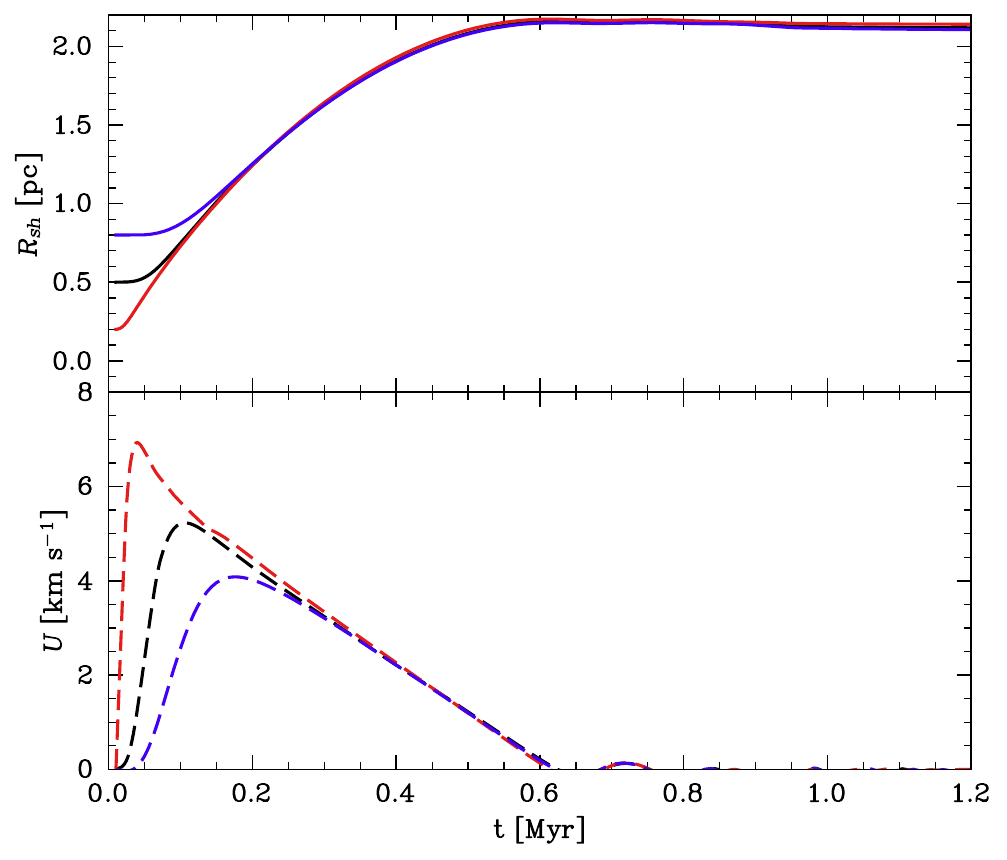}
\caption{Early evolution (shell radius and velocity in the upper and bottom panels, respectively) of the same model presented in section \ref{SFF_section} (see Fig. \ref{single_SSF}), but for three different initial values of the shell position, $r_0=0.2, 0.5$ and 0.8 pc. }
\label{plot_diff_r0}
\end{figure}

We nevertheless adopted a fixed initial radius because, in our setup (see Section~\ref{shell_evolution}), models initialized with slightly different shell positions rapidly converge to the same evolutionary track. Fig. \ref{plot_diff_r0} illustrates this feature for the same model presented in Fig. \ref{single_SSF} (section \ref{SFF_section}). The figure presents the early evolution of the shell radius (top panel) and velocity (bottom panel) for three different values of the starting position of the shell: $r_0 = 0.2, 0.5$ and 0.8 pc. The remaining input parameters are the same. By $t \sim 0.4$ Myr, all models converge to an indistinguishable solution. All other model variables show the same convergence behavior, and the subsequent evolution is insensitive to the choice of the initial shell position. 

For this reason, we decided to set $r_0 = 0.5$ pc as  fiducial initial condition for all models, while explicitly acknowledging that the thin-shell formation radius and time is not universal and varies with cloud properties in more realistic simulation setups.

\renewcommand\thesection{C}
\section{The star formation factory: existence of standing solutions}\label{standing_math}
\subsection{On the stability of the StSh models}
We now investigate the stability of StSh models, i.e., assuming a standing solution ($U=0$ and $\dot{U}=0$), we linearize the system of equations \ref{mod1}--\ref{energy_eq} around the equilibrium point. The perturbed equations for the shell radius, velocity, stellar and shell masses, are given by:
\begin{equation}
    \delta \dot{R}=\delta U,
\end{equation}
\begin{equation}
    \delta \dot{M_{\rm{sc}}}=-\frac{GM_{\rm{sh}}^2}{2 c_{\rm{s}}R^3} \delta R -\frac{3 M_{\rm{sh}}}{R} \delta U+\frac{GM_{\rm{sh}}}{2 c_s R^2} \delta M_{\rm{sh}},
\end{equation}
\begin{equation}
    \delta\dot{M_{\rm{sh}}}= 4 \pi R^2 \rho_{\rm{cl}} \delta U - \delta \dot{M_{\rm{sc}}},
\end{equation}

\begin{align}
M_{\rm{sh}} \delta \dot{U}= & \delta R \left[-4 \pi R P_{\rm{th}} 
+\frac{G}{R^3}\left(M_{\rm{sh}}^2+M_{\rm{sh}}M_{\rm{sc}} \right) \right]
 \notag \\
& +4 \pi R^2 \rho_{\rm{cl}}v_{\rm{cl}}\delta U -\frac{G\left(M_{\rm{sh}}+M_{\rm{sc}} \right)}{R^2} \delta M_{\rm{sh}}
\notag \\
& - \frac{GM_{\rm{sh}}}{R^2} \delta M_{\rm{sc}} +\frac{3 \left(\gamma-1 \right)}{R} \delta E_{\rm{th}}.
\label{deltaU}
\end{align}
Note that these equations make use of the star formation rate for a standing solution (see Equations \ref{cond_fragmentation}--\ref{star_formation_recipe}):
\begin{equation}
    \dot{M_{\rm{sc}}}=\frac{GM_{\rm{sh}}^2}{4c_{\rm{s}}R^2},
\end{equation}
and the equilibrium condition (see Equation \ref{mod1}):
\begin{equation}
    4 \pi R_{\rm{sh}}^2 \left(P_{\rm{th}}-\rho_{\rm{cl}}v_{\rm{cl}}^2\right)=F_{\rm{g}}.
\end{equation}
The linearized perturbed energy equation is:
\begin{equation}
\delta \dot{E_{\rm{th}}}= - 4 \pi R^2P_{\rm{th}} \delta U +\chi_{\rm{e}},
\end{equation}
where 
\begin{equation}
   \chi_{\rm{e}}\left( t\right)=L_{\rm{w}}-Q_{\rm{w}} 
\end{equation}
is a time-dependent, non-linear term, which implies that the system has an external input (non-homogeneous system). 

We simplify the system of equations and eliminate $M_{\rm{sc}}$ by making use of mass conservation. Therefore, we are interested in finding the stability conditions for:

\begin{equation} \label{fundamental_lin_eq}
    \mathbf{\dot{x }\left(t \right)} = \mathbf{J} \mathbf{x\left(t \right)} + \mathbf{b\left(t \right)},
\end{equation}
where $\mathbf{x}^{T}= \left[R, U, M_{\rm{sh}}, E_{\rm{th}} \right]$, $\mathbf{b}^{T}= \left[0, 0, 0, \chi_{\rm{e}}\left(t \right) \right]$, and:
\begin{equation}
\mathbf{J} =
\begin{bmatrix}
0 & 1 & 0 & 0 \\
A_1 & A_2 & A_3 & A_4 \\
B_1 & B_2 & B_3 & 0 \\
0 & C_2 & 0 & 0
\end{bmatrix},
\end{equation}
with entries:
\begin{equation}\label{Jac1}
    A_1M_{\rm{sh}} = 4 \pi R P_{\rm{th}}-8 \pi R \rho_{\rm{cl}} v_{\rm{cl}}^2-4 \pi G \rho_{\rm{cl}}M_{\rm{sh}},
\end{equation}

\begin{equation}
    A_2  = \frac{4 \pi R^2 \rho_{\rm{cl}}v_{\rm{cl}}}{M_{\rm{sh}}},
\end{equation}
\begin{equation}
    A_3  = -\frac{GM_{\rm{sc}}}{R^2M_{\rm{sh}}},
\end{equation}
\begin{equation}
    A_4  = \frac{3(\gamma-1)}{R M_{\rm{sh}}},
\end{equation}

\begin{equation}
    B_1 = \frac{GM_{\rm{sh}}^2}{2c_{\rm{s}}R^3},
\end{equation}
\begin{equation}
    B_2 = 4 \pi R^2 \rho_{\rm{cl}}+3 M_{\rm{sh}}/R,
\end{equation}
\begin{equation}
    B_3= -\frac{GM_{\rm{sh}}}{2 c_{\rm{s}}R^2},
\end{equation}
and 
\begin{equation}\label{Jac2}
    C_2 = - 4 \pi R^2  P_{\rm{th}}.
\end{equation}
\subsection{Stability of the homogeneous sub-system}
We first study the case of the homogeneous sub-system, i.e., Equation \ref{fundamental_lin_eq} with $\mathbf{b}=0$. The characteristic equation for the Jacobian $\mathbf{J}$ is:
\begin{equation}\label{characteristic_eq}
    \lambda^4+a\lambda^3+b\lambda^2+c\lambda=0,
\end{equation}
where:
\begin{equation}
    a= - \left(A_2+B_3 \right),
\end{equation}
\begin{equation}
    b = -A_1+A_2B_3-A_3B_2-A_4C_2, 
\end{equation}
and:
\begin{equation}
    c = A_1B_3-A_3B_1+A_4B_3C_2.
\end{equation}
Equation \ref{characteristic_eq} has a pole in $\lambda=0$, which implies that the system is not asymptotically stable. However, it can be marginally stable if the remaining poles have negative real parts. According to the Routh-Hurwitz theorem, this is the case if the coefficients of Equation \ref{characteristic_eq} satisfy $a,b,c>0$ and $ab>c$. From Equations \ref{Jac1}--\ref{Jac2}, these conditions are:
\begin{equation}
    \frac{-4 \pi R^2 v_{\rm{cl}}\rho_{\rm{cl}}}{M_{\rm{sh}}}+\frac{G M_{\rm{sh}}}{2 c_{\rm{s}}R^2}>0,
\end{equation}

\begin{align}
\frac{4 \pi R}{M_{\rm{sh}}} \left( P_{\rm{th}}+2 \rho_{\rm{cl}}v_{\rm{cl}}^2\right) & +4 \pi G \rho_{\rm{cl}} \left(1+M_{\rm{sc}}/M_{\rm{sh}} \right)
\notag \\
& +\frac{3GM_{\rm{sc}}}{R^3}-\frac{2\pi G \rho_{\rm{cl}}v_{\rm{cl}}}{c_{\rm{s}}}>0
\end{align}

\begin{align}
 G P_{\rm{th}}+2 G \rho_{\rm{cl}} v_{\rm{cl}}^2 & +\frac{ G^2 \rho_{\rm{cl}}M_{\rm{sh}}}{R}
\notag \\
& +\frac{G^2M_{\rm{sh}}M_{\rm{sc}}}{4 \pi R^4}>0
\end{align}

\begin{align}
 & -\left[ \left( P_{\rm{th}}+2 \rho_{\rm{cl}}v_{\rm{cl}}^2\right)\frac{4 \pi R}{M_{\rm{sh}}^2} +\frac{G^2M_{\rm{sh}}}{4c_{\rm{s}}R^4}\right]4 \pi R^2 \rho_{\rm{cl}}v_{\rm{cl}} 
\notag \\
& +\frac{2 \pi G^2 \rho_{\rm{cl}}M_{\rm{sc}}}{c_{\rm{s}} R^2}+\frac{G^2M_{\rm{sc}}M_{\rm{sh}}}{c_{\rm{s}}R^5}>0.
\end{align}
These inequalities always hold for our models in any case, given that $v_{\rm{cl}} < 0$ in our calculations. This implies that the remaining poles of Equation \ref{characteristic_eq} always have negative real parts. Thus, the homogeneous component of the system (given by Equation \ref{fundamental_lin_eq}) is always marginally stable. 

\subsection{On the stability of the complete system}
In the last section, it was shown that the homogeneous system has a 3D stable bundle and a center (or neutral) subspace (with $\lambda=0$). Rather than finding the complete solution, our interest is to establish what conditions on the driving term, $\mathbf{b}\left( t\right)$, lead to the stability of our model. For this reason, the focus will be on understanding the impact of $\mathbf{b}\left( t\right)$ on the neutral subspace. 

Let $\omega$ be the left null-vector of the Jacobian, i.e.,
    \begin{equation}
    \omega^{T} J =0, \hspace{0.2cm} \leftrightarrow \hspace{0.2cm} J^T \omega=0
    \end{equation}
    Solving for our system:
    \begin{equation}
        \omega^T= \left[ -C_2, 0,0,1\right].
    \end{equation}

Therefore, the projection of $\mathbf{b}\left( t\right)$ into the center mode is:
\begin{equation}    
    P \left( t\right)= \omega^{T} \cdot b\left( t\right)= \chi_e\left( t\right)
\end{equation}
Thus, the system is stable as long as:
\begin{equation} \label{con_stability_f}
    \lim_{\tau \to \infty} \frac{1}{\tau} \int_{t_0}^{t_0+\tau} \chi_e(s)\, ds = 0
\end{equation}
where $t_0$ is the onset of the equilibrium state.
\begin{figure}[htb]
\centering
	\includegraphics[width=\columnwidth]{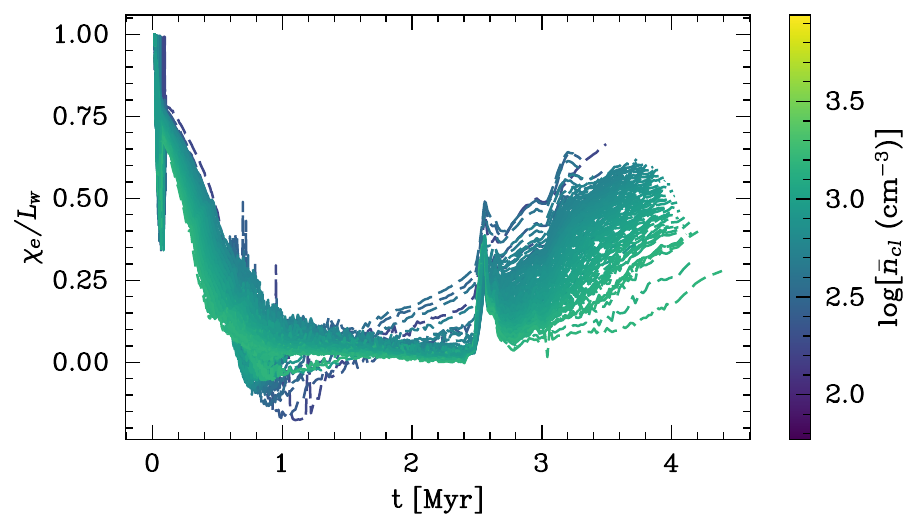}
    \caption{Time evolution of $\chi_e/L_{\rm{w}}$ for all the IZw18 StSh models, where the color bar presents the initial gas cloud number density.} 
\label{time_delta_IZw18}
\end{figure}

Thus, in physical terms, the stability analysis shows that a standing solution is stable as long as the cooling rate matches, in average, the increasing rate of the mechanical power. 

\begin{figure}[htb]
\centering
	\includegraphics[width=\columnwidth]{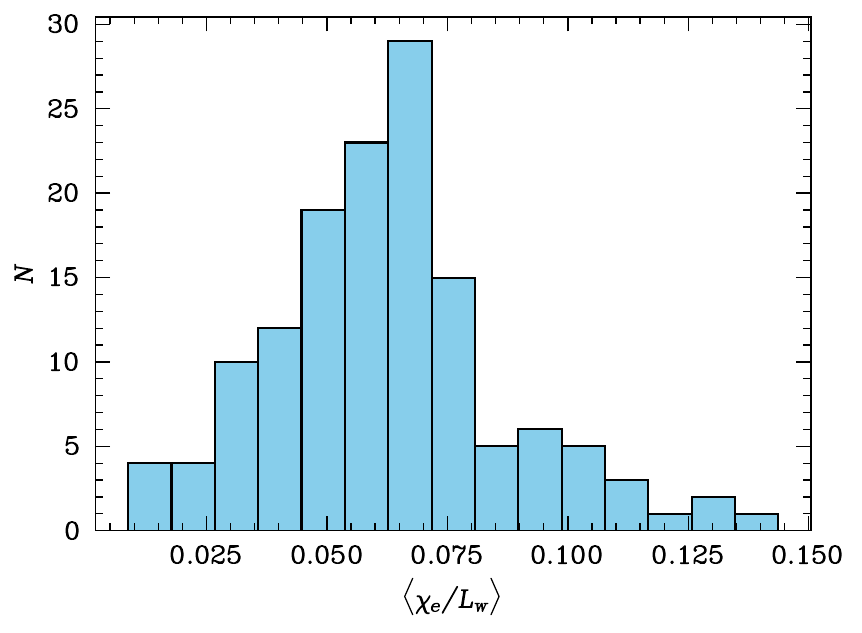}
    \caption{Histogram of the values of $\langle \chi_{\rm{e}} / L_{\rm{w}} \rangle$ obtained for the IZw18 StSh models, where the average is taken from the duration of the StSh phase. }
\label{hist_IZw18}
\end{figure}

We verified this analysis with our numerical calculations. Fig. \ref{time_delta_IZw18} presents the time evolution of $\chi_e/L_{\rm{w}}$ for all the IZW18 StSh models. Note how during the StSh phase, $\chi_e/L_{\rm{w}} \rightarrow 0$, which is expected from our previous  analysis. In addition, we also estimated the time average:
\begin{equation}
   \langle \chi_{\rm{e}} / L_{\rm{w}} \rangle = \frac{1}{\tau} \int_{t_0}^{t_0+\tau} \chi_{\rm{e}}(s) /L_{\rm{w}}(s) \, ds,
\end{equation}
where the integration was performed during the entire duration of the StSh stage ($\tau$). Fig. \ref{hist_IZw18} shows the results for the IZw18 case. The peak of the distribution is $\langle \chi_e / L_{\rm{w}} \rangle \sim 0.07$, which implies $Q_{\rm{w}} \gtrsim 0.9 L_{\rm{w}}$. Hence, all the StSh models are highly radiative, confirming that the condition given by Equation \ref{con_stability_f} is satisfied by the StSh models.

\renewcommand\thesection{D}
\section{On the uniqueness of the standing shell solutions}\label{star_formation_mode}
In the model setup of this work, once the core of the cloud has been swept-up by the shell, star formation in the core should stop and thus shell star formation should be dominant. However, we verified the robustness of the standing-shell solutions. Indeed, we performed an additional set of simulations for the IZw18 metallicity case in which shell star formation was completely suppressed, while core star formation was allowed to proceed continuously throughout the evolution. This setup differs from our fiducial models only in the star formation prescription. For this test,  we considered only a subset of our studied parameter space, with initial cloud masses in the range $5.0 \leq \log [M_{\rm{gas}} (M_{\odot})] \leq 5.3$.

Figure~\ref{single_SSF} presents the results of these new calculations (only StSh models are shown in this figure). Note how the stellar mass grows continuously throughout the entire evolution as given by the core star formation prescription. The figure also shows that standing shell solutions still develop under these conditions. Although the evolution is somewhat more oscillatory than in our main models, the qualitative behavior remains unchanged. This demonstrates that the existence of standing solutions is not a specific consequence of including star formation in the shell.

Instead, these solutions appear to arise from the dynamical balance between feedback, gravity, and the ram pressure of the collapsing cloud. This is further supported by previous results from our group \citep{2023MNRAS.521.5686K}, which showed similar standing solutions under substantially different modeling assumptions in collapsing clouds. Finally, note that changing the star formation prescription would indeed change the location of standing solutions in the parameter space composed by the initial gas cloud mass and radius (i.e., Fig. \ref{par_space_IZw18} and \ref{param_space_dwarfA}), but studying this in detail is outside the scope of the present work. 
\begin{figure}[htb]
\centering
	\includegraphics[width=0.95\columnwidth]{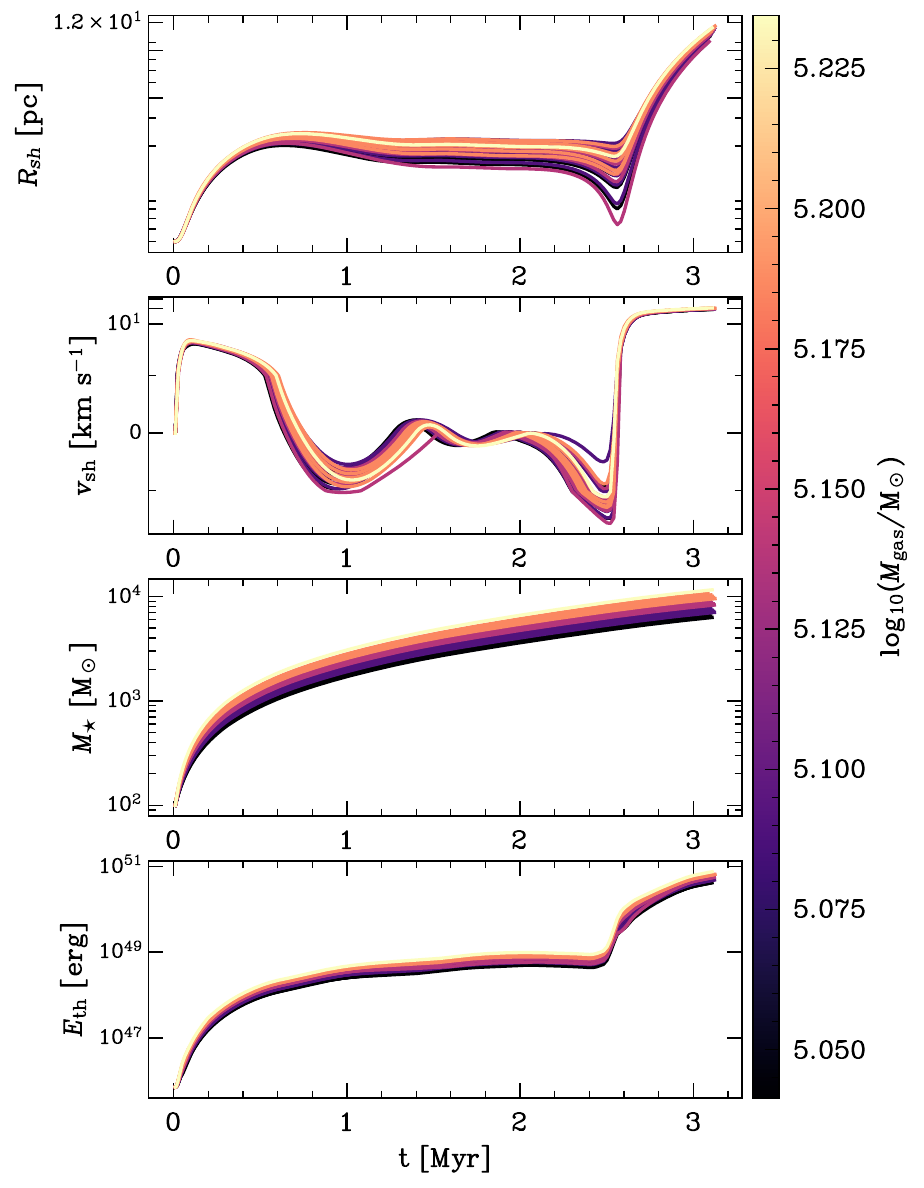}
    \caption{From top to bottom: shell radii and velocity, star cluster mass, and thermal energy of the bubble for a set of models with continuous core star formation and with no star formation in the shell. The color represents the log of the initial cloud mass for each model.}
\label{single_SSF}
\end{figure}

\renewcommand\thesection{E}
\section{Results of Model calculations}\label{plots_all_results}
\begin{figure*}[htb]
\centering
	\includegraphics[width=1.7\columnwidth]{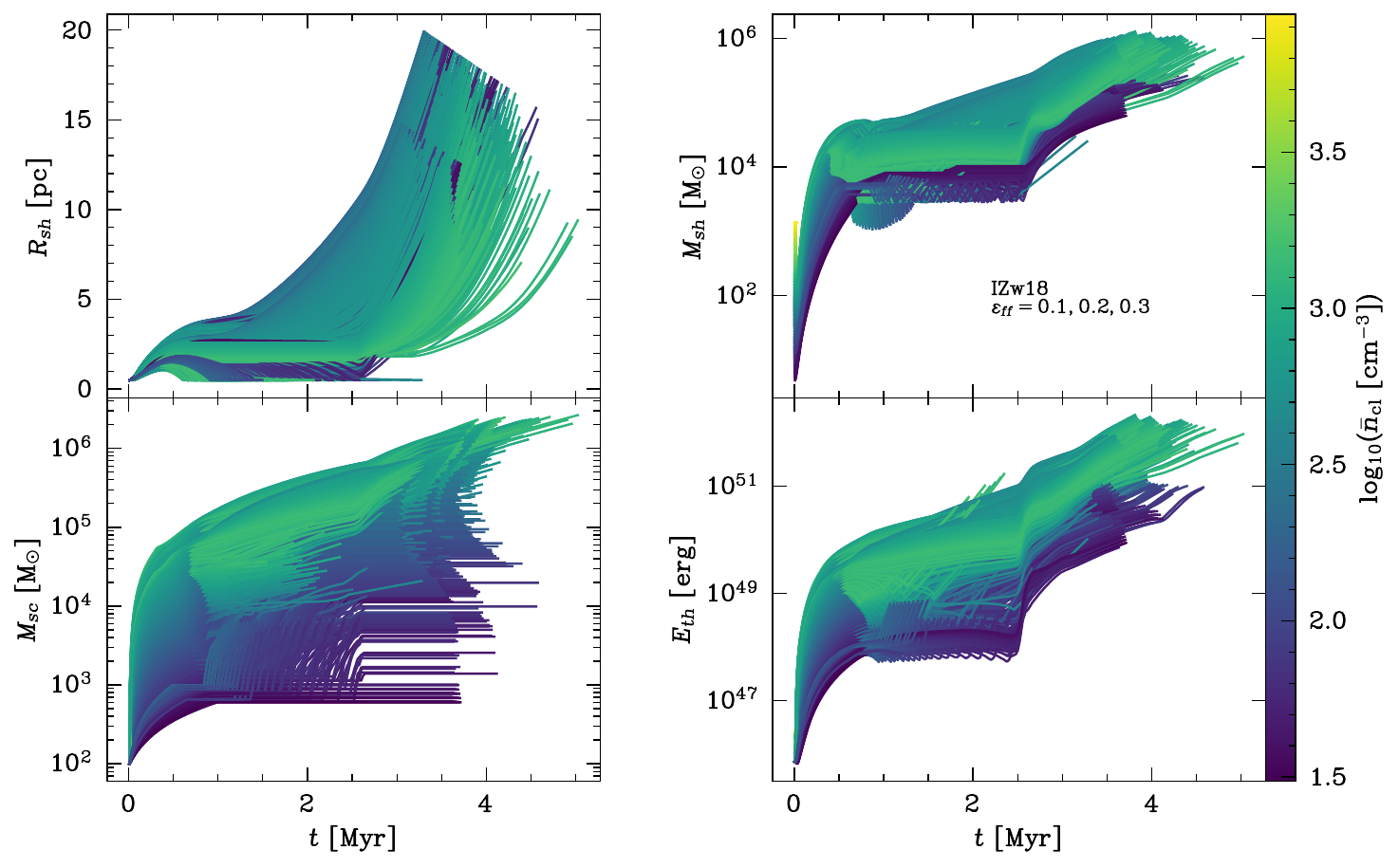}
\caption{Same as Fig. \ref{IZw18_all} but for the whole set of models.}
\label{IZw18_all_all}
\end{figure*}

\begin{figure*}[htb]
\centering
	\includegraphics[width=1.7\columnwidth]{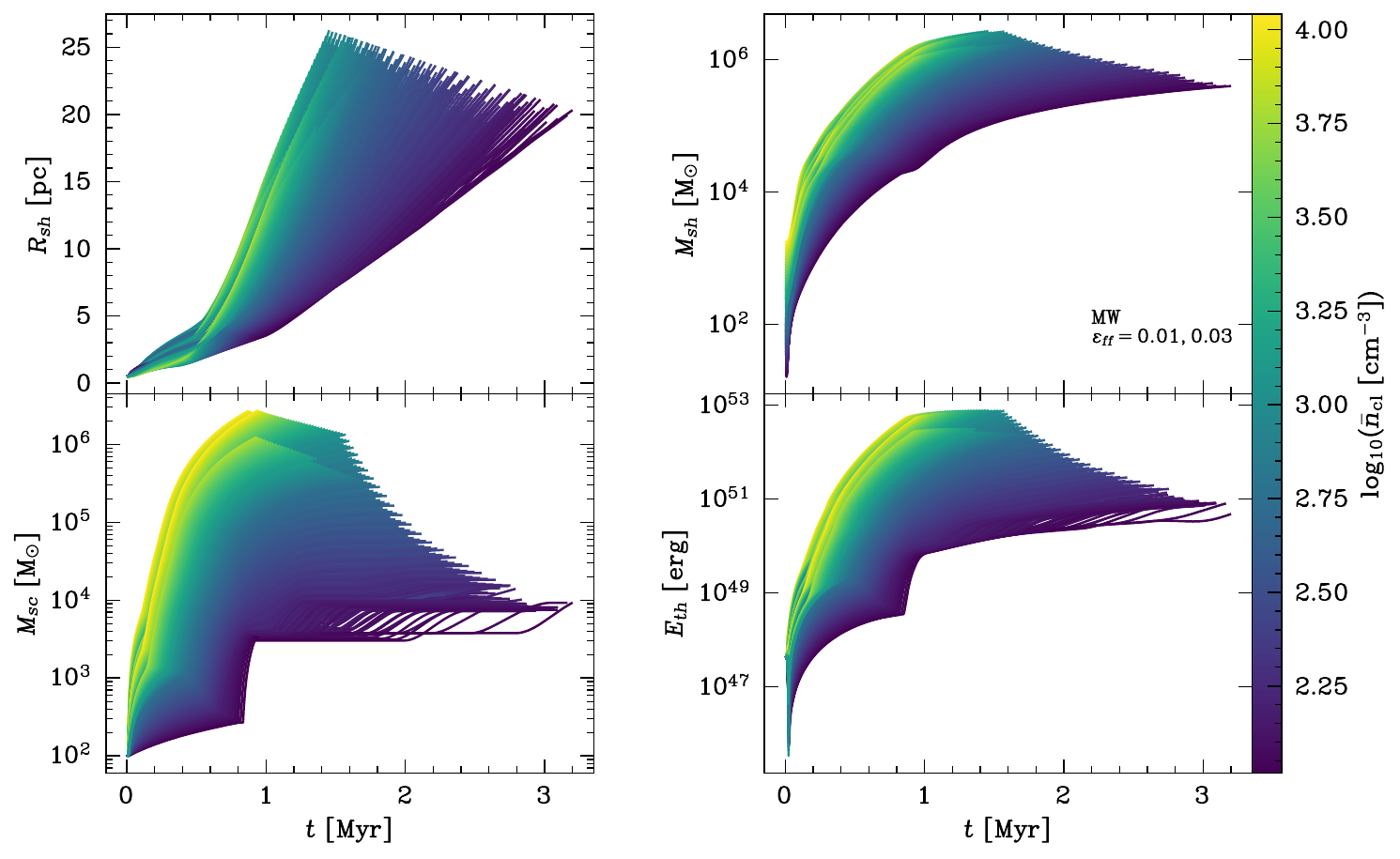}
\caption{Same as Fig. \ref{IZw18_all} but for the MW metallicity and the two values of $\epsilon_{\rm{ff}}$ considered for this case.}
\label{ap_model1}
\end{figure*}

\begin{figure*}[htb]
\centering
	\includegraphics[width=1.7\columnwidth]{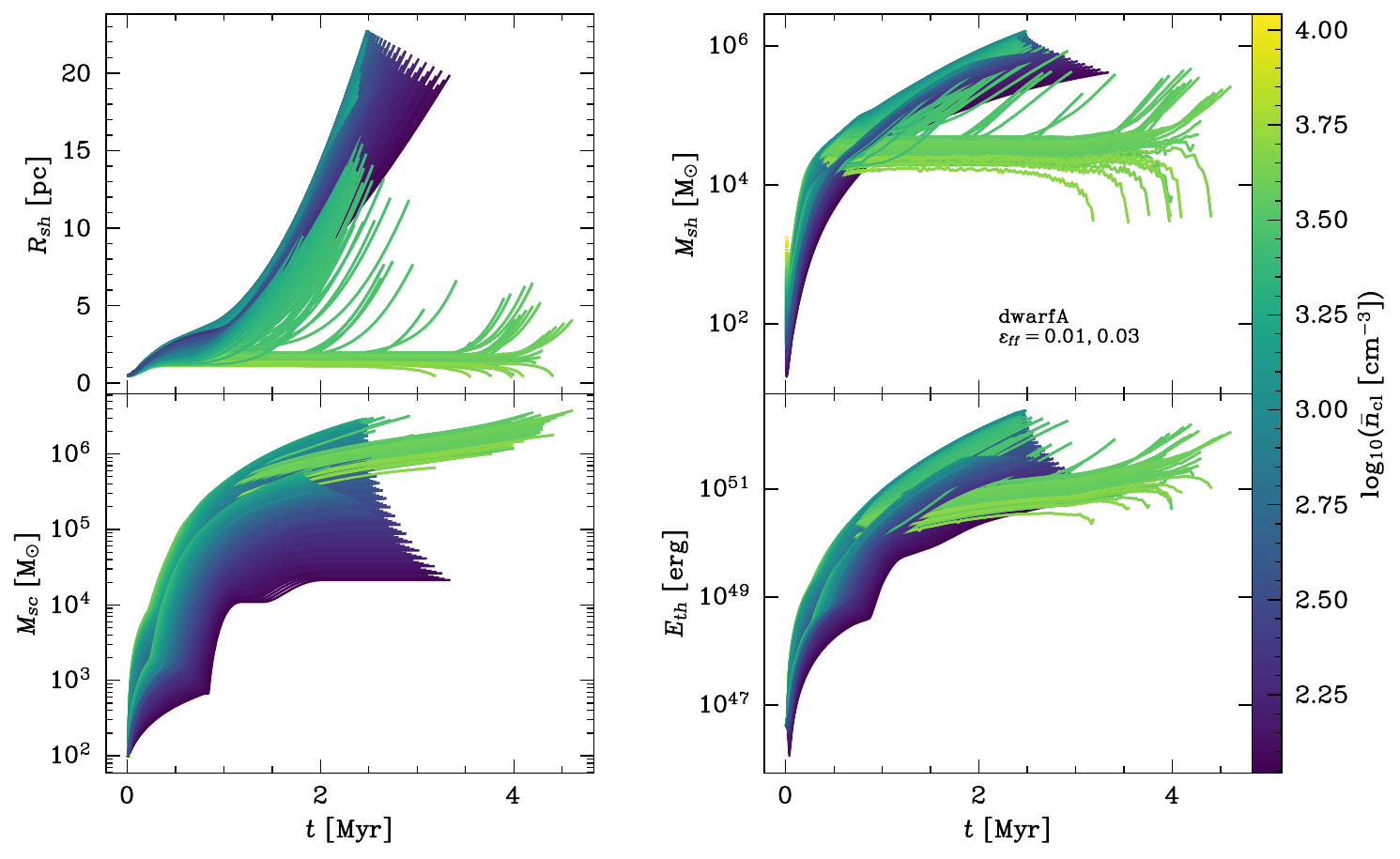}
\caption{Same as Fig. \ref{IZw18_all} but for the dwarfA metallicity and the two values of $\epsilon_{\rm{ff}}$ considered for this case.}
\label{ap_model2}
\end{figure*}

Figs. \ref{IZw18_all_all}, \ref{ap_model1} and \ref{ap_model2} present all the results for the IZw18, dwarfA and MW metallicity calculations, where the panels are the same as in Fig. \ref{IZw18_all}. 

\begin{figure*}[htb]
\centering
	\includegraphics[width=1.6\columnwidth]{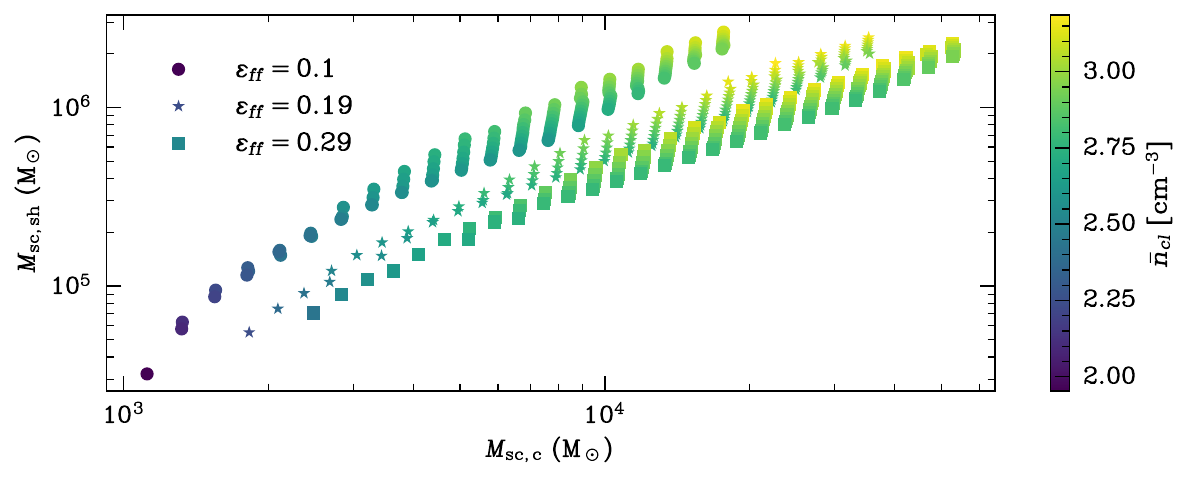}
\caption{The total mass ($M_{\rm{sc,c}}$) formed from core star formation (Equation \ref{core_SFR}) and the total mass ($M_{\rm{sc,sh}}$) from shell star formation (Equation \ref{cond_fragmentation}) for the StSh models of Fig. \ref{par_space_IZw18}, with the color bar presenting the average initial gas number densities of the clouds. The symbols present different values of $\epsilon_{\rm{ff}}$ (see the inset). }
\label{eglobal_IZw18b}

\end{figure*}

Fig. \ref{eglobal_IZw18b} presents a comparison between the total stellar mass ($M_{\rm{sc,c}}$) that is formed from core star formation over a free-fall time (Equation \ref{core_SFR}), and the mass ($M_{\rm{sc,sh}}$) formed from shell star formation when the shell becomes gravitationally unstable (Equation \ref{cond_fragmentation}). The symbols in this plot represent different values for $\epsilon_{ff}$ (see the figure inset) while the color bar shows the initial gas cloud number density. Only StSh models from Fig. \ref{par_space_IZw18} are shown in this figure (i.e., IZw18 models).  As expected, $M_{\rm{sc,c}}$ and $M_{\rm{sc,sh}}$ are correlated given that they both depend on the gas density. Nevertheless, note that the shell star formation is almost two orders of magnitude larger than the mass formed from the core for most of the models. 

\end{appendix}

\end{document}